\documentclass[journal=jacsat,manuscript=article]{achemso}

\usepackage[version=3]{mhchem} 
\usepackage{siunitx}
\usepackage{geometry}
\author{Michael Verhage}
\affiliation{Molecular Materials and Nanosystems - Department of Applied Physics, Eindhoven University of Technology, The Netherlands}
\author{Pantelis Bampoulis}
\affiliation{Faculty of Science and Technology, Physics of Interfaces and Nanomaterials, University of Twente, The Netherlands}
\author{Marco D. Preuss}
\affiliation{Institute for Complex Molecular Systems (ICMS), Eindhoven University of Technology, PO Box 513, 5600 MB, Eindhoven, The Netherlands}
\alsoaffiliation{Laboratory of Macromolecular and Organic Chemistry,
Eindhoven University of Technology, The Netherlands}
\author{Ivo Filot}
\affiliation{Inorganic Materials $\&$ Catalysis - Department of Chemical Engineering and Chemistry, Eindhoven University of Technology, The Netherlands}
\author{Heiner Friedrich}
\affiliation{Laboratory of Physical Chemistry, Department of Chemical Engineering, Eindhoven University of Technology, 5600 MB Eindhoven, The Netherlands}
\alsoaffiliation{Institute for Complex Molecular Systems (ICMS), Eindhoven University of Technology, PO Box 513, 5600 MB, Eindhoven, The Netherlands}
\author{Rick R.M. Joosten}
\affiliation{Laboratory of Physical Chemistry, Department of Chemical Engineering, Eindhoven University of Technology, 5600 MB Eindhoven, The Netherlands}
\alsoaffiliation{Center for Multiscale Electron Microscopy, Department of Chemical Engineering, Eindhoven University of Technology, 5600 MB Eindhoven, The Netherlands}
\author{E.W. Meijer}
\affiliation{Institute for Complex Molecular Systems (ICMS), Eindhoven University of Technology, PO Box 513, 5600 MB, Eindhoven, The Netherlands}
\alsoaffiliation{Laboratory of Macromolecular and Organic Chemistry,
Eindhoven University of Technology, The Netherlands}
\author{Kees Flipse}
\affiliation{Molecular Materials and Nanosystems - Department of Applied Physics, Eindhoven University of Technology, The Netherlands}
\email{c.f.flipse@tue.nl}

\title
  {Charge transport modulation by a redox supramolecular spin-filtering chiral crystal}


\begin{document}

\begin{abstract}
  The chirality induced spin selectivity (CISS) effect is a fascinating phenomena correlating molecular structure with electron spin-polarisation in excited state measurements. Experimental procedures to quantify the spin-filtering magnitude relies generally on averaging data sets, especially those from magnetic field dependent conductive-AFM. We investigate the underlying observed disorder in the IV spectra and the origin of spikes superimposed. We demonstrate and explain that a dynamic, voltage sweep rate dependent, phenomena can give rise to complex IV curves for chiral crystals of coronene bisimide. The redox group, able to capture localized charge states, acts as an “impurity” state interfering with a continuum, giving rise to Fano resonances. We introduce a novel mechanism for the dynamic transport which might also provide insight into the role of spin-polarization. Crucially, interference between charge localisation and delocalisation during transport may be important properties into understanding the CISS phenomena. 
\end{abstract}


\section{Introduction}
Heightened research into the field of chiral supramolecular self-assembly by using non-convalent interactions has led to emergence of non-trivial electron transport phenomena through metal-molecule junctions \cite{Waldeck2021TheMaterials, Naaman2022NewMolecules, Kulkarni2020a}. Of fundamental and application driven importance is the intricate interplay between molecular chirality and spin-polarisation (SP) of electrons as is broadly observed in electron transmission \cite{Gohler2011SpinDNA}, magnetoresistance devices \cite{Sang2021TemperatureAzurin, Michaeli2017AMagnets, Mondal2021SpinSolvents, Lu2019Spin-dependentPerovskites} and electrochemistry \cite{Bhowmick2022Spin-inducedElectropolymerization}. From an experimental point of view, correlating molecular structure with SP property in achieving a deeper understanding of the mechanism behind CISS is important, as it would allow uncovering molecular design rules. Designing specific molecules with an active control over the spin-polarisation combined with low electrical resistance for high current densities should hence become possible. Evidence is emerging that besides chirality, the molecular layer thickness \cite{Mondal2021SpinSolvents} and polarisation\cite{Naaman2022NewMolecules}, are properties regulating the CISS effect. Although considerable research \cite{Waldeck2021TheMaterials} has been
devoted to demonstrate the link between chirality and spin filtering and forming theoretical models of the CISS effects, insights into the electronic transfer mechanisms and modulation of charge through a molecular transport junction are still to be gained. Our stacked crystalline layers are distinct from a tunneling junction \cite{Nozaki2013AJunctions} by a layer thickness exceeding that of a single molecule, and hence charge transfer by tunneling is not expected to be the dominant transport mechanism  \cite{Xu2005LargeTransistor, Chen2007MeasurementConductance, McCreery2009ElectronJunctions, He2003DiscreteWires, Jia2020RedoxJunctions}

The aim of this work is to correlate two fundamental properties of a chiral molecule possessing two imide redox groups \cite{Kulkarni2020a} and an electronic conductive continuum formed by the $\pi$ - $\pi$ stacked coronene core or by a highly diffuse molecular orbital (HDMO) \cite{Bohl2017Low-lyingAlkanes}; introducing significant charge delocalisation and giving rise to the CISS effect via Fano resonances. The combination of these two properties and the proposal of a charge modulation model \cite{Migliore2013IrreversibilityJunctions}, are what we argue are a missing link in the understanding of the CISS effect of such molecules. Based on conduction measurements we suggest a novel mechanism for a microscopic understanding of the measured charge dynamic effects. Charge localisation, in our case driven by imide groups \cite{Li2015Three-StateTuning} and side groups, are able to induce localized charge states which acts as an “impurity” state, similar to Andersons model \cite{Anderson1966TheoryMetals} \textbf{(Fig. \ref{fig:AFM_Crystals}A)}. To this end, we study chiral coronene bisimide (CBI-GCH) which demonstrated self-assembly into P- or M-type second order hierarchical helices as previously reported by Kulkarni \textit{et al.} \cite{Kulkarni2013Self-assemblyAmplification} and yielded very high SP degree up to 50\% at room temperature \cite{Kulkarni2020a}. Such significant SP at room temperature would make SP based devices a possibility and is highly desirable for example in catalytic applications \cite{Sang2022ChiralityReduction, Liang2022EnhancementElectrodes, Mtangi2017ControlSplitting}. The charge transport of such redox molecular junction is described by Migliore and Nitzan \cite{Migliore2013IrreversibilityJunctions}  within transition rate process models. Indicative for the fast increase of the bias voltage are what we assign as Fano like peaks (\textbf{Fig. \ref{fig:AFM_Crystals}B}) and (\textbf{Fig. \ref{fig:AFM_Crystals}C}). Fano resonances emerge naturally from the coupling of a discrete state with an electronic continuum \cite{Nozaki2013AJunctions}. Similar coupled systems relying on quantum interference, have been created by the coupling of molecular magnets to a continuum introduced by carbon nanotubes, yielding spin polarisation based on Fano resonances \cite{Urdampilleta2011SupramolecularValves, Hong2013Fano-resonance-drivenMagnets}

One robust experimental approach of identifying the transmission of the preferential spin of a chiral molecule is derived from averaging many IV spectra and observing a change in current-voltage e.g. resistance, between a two-state magnetic field. A consequence of this approach is that, by approximation, a relatively "smooth" exponential IV curve is observed. Yet, even after averaging many IV spectra, the dynamics in the averaged IV curves is not suppressed \cite{Nguyen2022CooperativeMicroscopy, Bhowmick2022Spin-inducedElectropolymerization, Ko2022TwistedTemperature}. From single IV spectra, observation of dynamic charge transfer is expressed in "spikes" \cite{Migliore2013IrreversibilityJunctions}: a rapid rise and fall of the current over a small change in electric field magnitude \cite{Kulkarni2020a, Ko2022TwistedTemperature, Mishra2020SpinPolymers, Mondal2020Long-RangeMagnetization, Lu2021HighlyHalides}. We study CB-GCH with conduction-AFM (C-AFM) and discover charge transport variability expressed in non-monotonous IV curves and sequential IV sweeping induced reduction in electrical resistance. The transport characteristics are very dependent on lab-accessible parameters such as voltage sweep rate, normal force of the C-AFM probe and the local morphology of the chiral aggregate. Finally, we discuss molecule properties we predict can be actively be tuned to enhance the CISS magnitude and enhance current densities. 

\section{Results and Discussion}
\subsubsection{Structural characterisation of CBI-GCH supramolecular crystals}
\begin{figure*}
    \centering
       \includegraphics[scale=0.37]{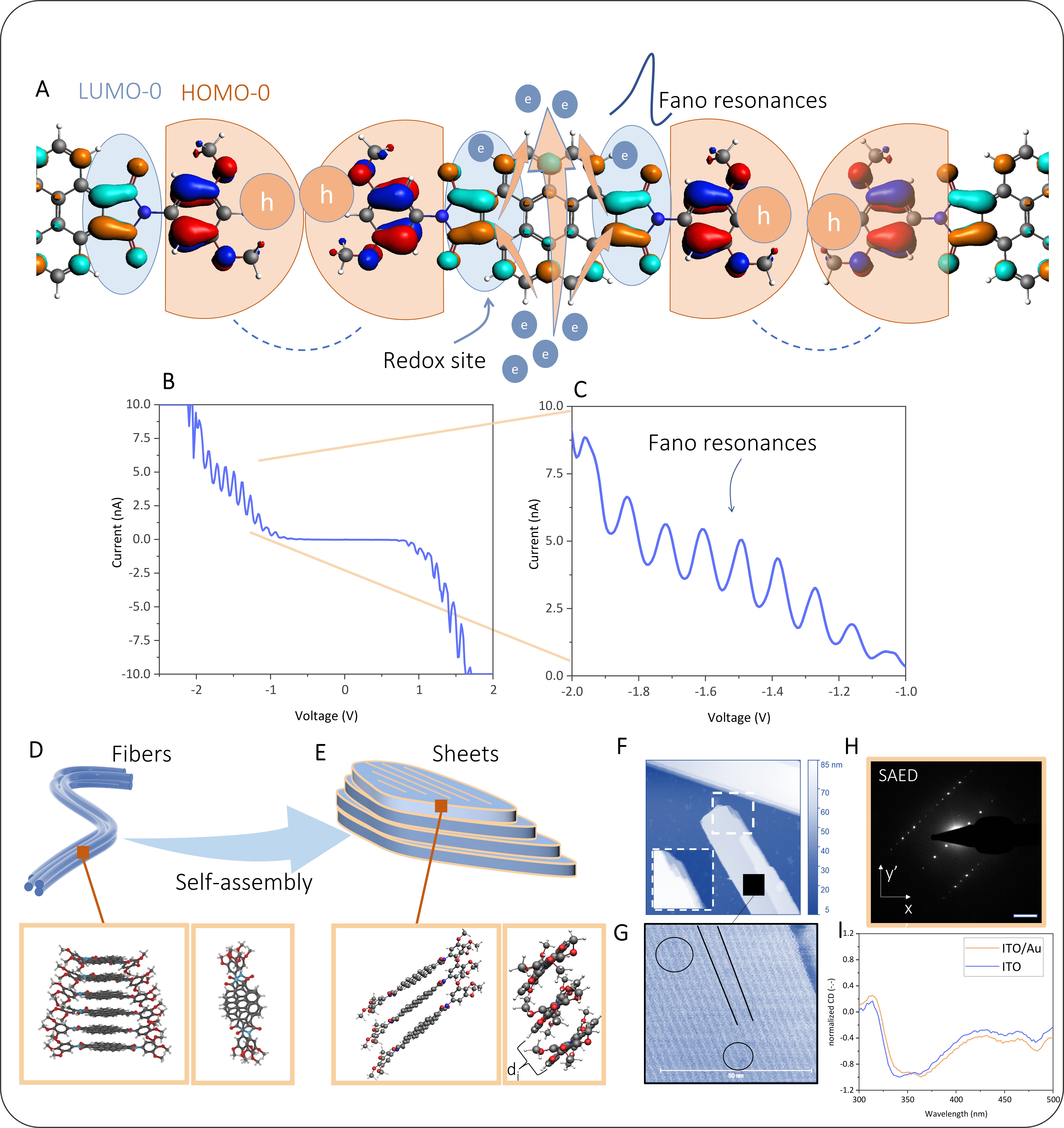}
            \caption{\textbf{Charge modulation within CBI-GCH crystals} 
            (\textbf{A}) Crystalline CBI-GCH depicted with imide redox groups with LUMO-0 (blue oval) and HOMO-0 (red semi-circle) located at the side groups, trapping charge and modulating the main transport channel \cite{Migliore2013IrreversibilityJunctions}. The coupling between the localized charges and a continuum across the coronene core results in Fano resonances. (\textbf{B, C}) Fano resonances observed in IV spectra. (\textbf{D}) CBI-GCH helical fibers with DFT optimized geometry. (\textbf{E}) Crystallisation of CBI-GCH in MCH with DFT optimized geometry. (\textbf{F}) AFM topography taken at the terminus of a crystal showing the stacked platelets. (\textbf{G}) High resolution AFM image indicating parallel fibers in the direction of the black dotted lines. The black circles indicate terminus of single supramolecular polymers. (\textbf{H}) Selected area diffraction (SAED) of a CBI-GCH crystal. (\textbf{I}) CD spectra of crystals deposited on ITO and ITO/Au substrates} 
                \label{fig:AFM_Crystals}
\end{figure*}

In our aim of studying the dynamics of CBI-GCH's redox induced charge modulation \cite{Migliore2013IrreversibilityJunctions} and consequential correlation to the CISS effect, we explored self-assembled, hierarchical ordering \cite{Yuan2019HierarchicallyCrystals} of CBI-GCH crystals. These crystals show polymorphism \cite{Yuan2019HierarchicallyCrystals} by yielding high quality crystals \textbf{(Fig. \ref{fig:AFM_Crystals}E, F)}. The conductive insights gained from well ordered crystals, enable further insights into the behaviour of the ductile supramolecular polymeric fibers \cite{Kulkarni2020a} and hence we regard the former as an enticing model system to study the CISS effect. 

The CBI-GCH discotics are self-assembled in methylcyclohexane (MCH) into helical aggregates driven by non-covalent interactions \cite{Kulkarni2020a}. Our CD (\textbf{Supp1}) and DFT geometry optimization \textbf{(Fig. \ref{fig:AFM_Crystals}D)} of this aggregate support helical self-assembly with a mean interdiscotic distance of \SI{3.5}{\angstrom} and a planar rotation of \SI{10.5}{\degree} \textbf{(Fig. \ref{fig:AFM_Crystals}a)}. Remarkable, the polymers self-assemble into stacked sheets \textbf{(Fig. \ref{fig:AFM_Crystals}B)} upon a relative short time scale of weeks as dissolved in MCH. For structural and conductive studies, these crystals were drop-casted (10 - \SI{20}{\micro\liter}, \SI{20}{\milli\mol}) onto several substrates of freshly cleaved mica and HOPG and a Si wafer. Across these substrates the crystal density widely varies, driven by isodesmic growth \cite{Kulkarni2013Self-assemblyAmplification} and drying induced concentration gradients. Elongated structures were observed with AFM \textbf{(Fig. \ref{fig:AFM_Crystals}F)}. We note that the crystals show clustering behaviour (\textbf{SuppS2}) which likely point to long range interaction, such as dipole driven electrostatic interactions \cite{Yuan2019HierarchicallyCrystals}. From AFM imaging, crystals with a mean length of several hundred nanometer were most commonly observed, but crystalline sheets up to \SI{15}{\micro\meter} wide (\textbf{SuppS3}) were also noted. Besides the lateral dimensions, the remarkable quality of crystallisation is supported by observing extremely smooth surfaces of the sheets, with a roughness below \SI{0.2}{\nano\meter}. Such smoothness rivals that of typical self-assembled monolayers, yet in our case crystals can be stacked of hundreds of molecular sheets. 

We examined the nanoscale ordering of several crystals with PeakForce tapping AFM and observe that the sheets consist out of parallel ordered supramolecular polymers (\textbf{Fig. \ref{fig:AFM_Crystals}D}), with a peak-to-peak spacing of $S$  = $\SI{2.5}{}\pm\SI{0.2}{\nano\meter}$. Because the diameter of CBI-GCH between the two peripheral benzene groups is around \SI{2.03}{\nano\meter}, it suggests alkyl chains are not fully extended and therefore potentially back-folded or interpenetrate amorphous in the sheets. We did observe significant adhesion between the AFM tip and the crystal surface, likely indicating the alkyl groups were sticking-out of the surface and adhering to the AFM tip (\textbf{SuppS4}). Supramolecular polymers are found to be of finite length as highlighted in the black circles (\textbf{Fig. \ref{fig:AFM_Crystals}G}) showing segmented supramolecular polymers.  This observation supports to notion that supramolecular polymers with a distribution in length crystallise in solution following the isodesmic growth \cite{Kulkarni2013Self-assemblyAmplification}. This degree of close packed crystallisation may be extremely beneficial for device application, ie magnetoresistance devices, compared to fibers which suffer from pin-holes \cite{Kulkarni2020a}.  

To gain further insight into CBI-GCH crystalline ordering, we turned to selected area diffraction (SEAD) with a transmission electron microscope (TEM). \textbf{Fig. \ref{fig:AFM_Crystals}H} shows a SAED pattern on a crystal (\textbf{SuppS5}). The sharp contrast of the diffraction peaks highlight the isotropic ordering within the crystals. The fiber-to-fiber spacing $S$ of \SI{2.49}{\nano\meter} is obtained. This matches the value of \SI{2.5}{\nano\meter} obtained with the high-resolution AFM very well. The intramolecular spacing is found to be \SI{4.7}{\angstrom}. To investigate whether the helical structure of the polymers was still preserved when crystallised, we simulated several crystalline ordering phases and simulated the frequency spectrum with a Fast Fourier Transform (FFT), to compare with the SAED pattern.  Because of the symmetrical shifted diffraction pattern, as a likely internal structure candidate we simulated an herringbone structure (\textbf{SuppS5}). Coronene is also known to crystallise in this phase \cite{Potticary2016AnGrowth}. The herringbone phase seems to best simulate the SAED pattern and the FFT is given in (\textbf{SuppS5}). This observation is a different assembly to the helical phase of the 1D polymers, highlighting polymorphic behaviour of CBI-GCH between an H-aggregate and a J-aggregate. Such behaviour is perhaps surprising and shows the strong dynamics of these supramolecular coronene based molecules. DFT calculations corroborate the polymorphic aggregation, with the J-aggregate found with same energy minima as the helix (\textbf{Fig. \ref{fig:AFM_Crystals}E}). The monomers are not rotated, but slipped in-plane, the interdiscotic distance is found to be \SI{4.5}{\angstrom}, which matches those obtained by SAED rather well. Future Molecular Dynamics calculations could possible shed further light into the phase transition dynamics \cite{Kulkarni2017SolventPockets}. Finally, we measured CD spectra of the crystals deposited on both ITO/Au and bare ITO substrates (\textbf{Fig. \ref{fig:AFM_Crystals}I}), and identified a chiral signal. Although we cannot exclude the presence of 1D helical fibers mixing with the signal, the CD signal is starkly different from those obtained from 1D fibers in solution \cite{Kulkarni2013Self-assemblyAmplification}. Hence, the ordering of CBI-GCH into crystals are likely to preserve the chirality, dictated by the point chirality embedded in the alkyl chains \cite{Kulkarni2020a, Kulkarni2013Self-assemblyAmplification}. 

\subsubsection{Electrical transport}

\begin{figure*}
    \centering
        \includegraphics[scale=0.31]{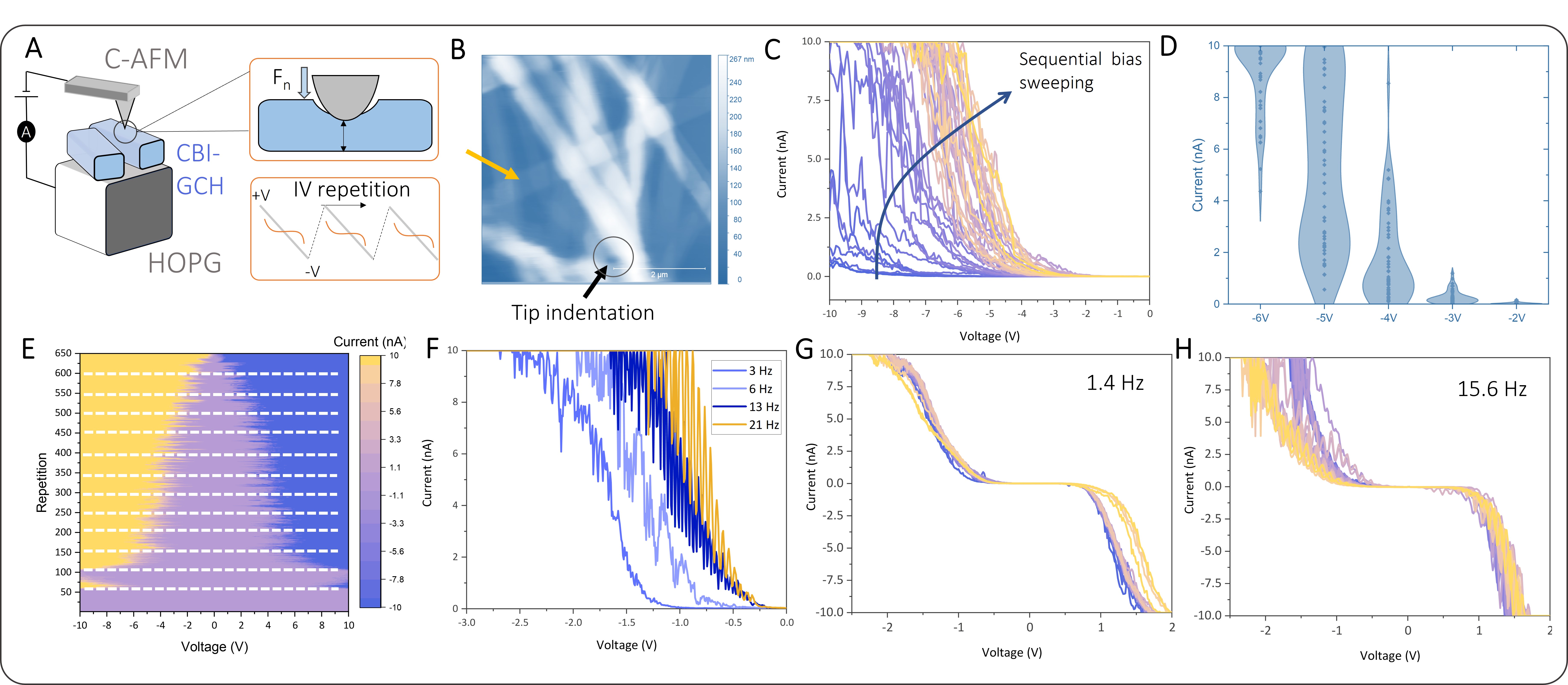}
            \caption{\textbf{C-AFM IV spectroscopy.} 
            (\textbf{A}) C-AFM experimental setup, highlighting the normal force $F_n$ of the tip used to control indentation of the tip into the crystals, and the reduction of layer thickness indicated with the black arrows. Sequential IV sweeps are schematically illustrated as swept between positive and negative bias voltage. (\textbf{B}) Topographic AFM image showing the location of spectroscopy taken at the yellow arrow. The black area indicates an area after large force tip indention. (\textbf{C}) \SI{20}{} sequential IV sweeps at the same location, with constant $F_n$, indicating gradual reduction of electronic gap. (\textbf{D}) Plotting the IV curves of \textbf{c} on a violin plot, shows non-normal distribution especially for larger bias voltages. (\textbf{E}) Gradually increasing $F_n$, for a \SI{40}{\nano\meter} thick crystal, by \SI{5}{\nano\newton} steps (white dashed lines) a gradual decrease of the gap can be observant, also for constant $F_n$ (between two white dashed lines) the gap gradually decreases. The colour plot shows \SI{650}{} IV curves taken. (\textbf{F})}
               \label{fig:C-AFM}
\end{figure*}

To shed light on the charge transport characteristics we turn to C-AFM (\textbf{Fig. \ref{fig:C-AFM}A}). The C-AFM tip acts as a movable electrode and the current is measured perpendicular to the surface. Crystals were first imaged with AFM, where care was taken to minimize lateral force induced damage (\textbf{SuppS6}). The crystals were drop-casted on HOPG and the local crystalline morphology and thickness (\textbf{Fig. \ref{fig:C-AFM}B}) was registered before commencing IV spectroscopy. An example of a \SI{20}{\nano\meter} crystal where IV spectroscopy was performed is highlighted with the yellow arrow in \textbf{Fig. \ref{fig:C-AFM}B}. Following, the tip was positioned on a location with the tip softly in repulsive force, at the snap-to-contact point, with a nominal normal force of \SI{12}{\nano\newton}. In \textbf{SuppS7} the force calibration is given. Furthermore, we choose to use diamond tips due to their extreme mechanical stability (\textbf{SuppS8}).  Insights into the dynamic memory effect \cite{Migliore2013IrreversibilityJunctions} and transport variation can be investigated by sequentially sweeping the bias between positive and negative voltages (\textbf{Fig. \ref{fig:C-AFM}A}) at the same aggregate location. For all IV spectra the ramp rate was prior set and defined the time taken for each voltage sweep. 

In \textbf{ Fig. \ref{fig:C-AFM}C} \SI{20}{} sequential IV sweeps taken at negative bias voltages, are superimposed which show unexpected dynamic resistance variation. The first few IV curves (dark blue) indicate the crystal layer is a strong insulator, with little current flowing even at bias as high as \SI{8}{\volt}. Gradually, by continuous sequential bias sweeping, the electrical gap is reduced. For the last \SI{5}{} curves (yellow curves), the gap stabilises around $\pm$\SI{2}{\volt}. We note that the tip force was kept constant during measurement, and no lateral drift was observed. The surprising characteristic of this dynamic gap reduction is it's reproducibility. Having observed over \SI{1000}{} IV spectra, the dynamic change in gap is steadily observed across many crystals of variable thickness and composition. Hence, we establish this behaviour as an intrinsic property of this crystalline aggregate. The large variability and spread in current also presents a problem for averaging the sequential IV spectra. In \textbf{Fig. \ref{fig:C-AFM}D}, violin plots of the current distribution are shown for several bias voltages. The non-normal distribution of the current is evident, which limits single location, repeated bias sweep averaging. For large bias, i.e. above \SI{6}{\volt}, saturation of the current by the current amplifier leads a skewed distribution. 

Following, we increased the normal force $F_n$ applied by the tip to the crystals step-wise by \SI{5}{\nano\newton} to reduce the layer thickness, as schematically illustrated in \textbf{Fig. \ref{fig:C-AFM}A}. The bias was swept \SI{50}{} times before increasing $F_n$ further. By reducing layer thickness the electrical resistance is reduced as transport becomes more efficient. We confirm tip indentation by observing on many occasion tip holes in the aggregate layer, an example is highlighted with the black circle (\textbf{Fig. \ref{fig:C-AFM}B}).    From this experiment (\textbf{Fig. \ref{fig:C-AFM}E}) the reduction of the gap is continuously observed by increasing $F_n$. Hence, the transport is not modulated by tunneling, as is also not to be expected for a layer thickness of \SI{20}{\nano\meter}, for we would have to observed an exponential reduction in gap resistance. The gradual gap decrease we also attribute to the highly ordered nature CBI-GCH molecules in crystalline phase, where the increased $F_n$ is not likely to distort significantly the lateral ordering which may be of case for the 1D fibers.

An important expression, and perhaps overlooked property, of the dynamic conductive behaviour of chiral molecular aggregates can be found in the peculiar rapid increase and decrease in current magnitude, we refer to as spikes or Fano resonances. In \textbf{Fig. \ref{fig:C-AFM}F} examples of such peak structure is given. The bias sweep rate was varied between \SI{3}{\hertz} and \SI{21}{\hertz}. These spikes are superimposed on the current transport and show a steep slope, indicative of highly efficient charge transport, within a small bias range of several tens of mV. The notion of spikes is reproducible throughout all IV spectra (over 1000 curves have been taken across many positions on the aggregates).  Hence, artificially removing the spikes by averaging many IV spectra across several locations, as predominately done in CISS literature \cite{Kulkarni2020a, Ko2022TwistedTemperature, Mishra2020SpinPolymers, Mondal2020Long-RangeMagnetization, Lu2021HighlyHalides}, a vital element of the charge transport is likely lost. For significant bias voltage exceeding several volts, (note the electric field is large, in the order of \SI{e8}{\volt\per\meter}), (\textbf{SuppS9}), a rapid increase of the current is observed. This regime always coincides with the emergence of peaks superimposed on the current transport, which highlights that the current fluctuations are not merely dependent on the bias potential but on the occurrence of significant charge flow \cite{He2003DiscreteWires}. The occurrence of peaks or Fano resonances, suggest that the CBI-GCH molecules switch their conductive state \textit{collectively}, as hundreds of molecules are simultaneously probed during C-AFM.

Dielectric breakdown can likely be excluded as the cause of the rapid increase in current, or spikes, because even after several sweeps the spikes in the current does not saturate.  Even a current decrease can be observed for larger voltages after several consecutive current spikes. For very thin layers, i.e. monolayers, and high voltages in excess of several volts, dielectric breakdown cannot be ruled out, but is only likely for thin crystalline layers and very high electric fields. Also, slightly increasing the tip force $F_n$, effectively reducing the junction thickness, seems not to lead to an electrical short, which cannot be explained by dielectric breakdown. Note the AFM tip force variation is small because of the active feedback working during the acquisition of the IV spectra. 

We consider the dynamic change in resistance by charge capture in light of the presence of imide redox sites \cite{Daub2021Imide-BasedBatteries} of CBI-GCH. The theoretical work of Migliore and Nitzan \cite{Migliore2013IrreversibilityJunctions} describes a model for a molecular junction containing redox sites. We conjecture the imide groups on CBI-GCH function as mediator of the transferring charge through the delocalized orbitals across the core of the molecule, in this case coronene. This enables the manifestation of interplay between charge transfer with different timescales \cite{Migliore2013IrreversibilityJunctions}. 
The presence of (noise) spikes for this molecule, and to extend more general in C-AFM CISS literature \cite{Kulkarni2020a, Ko2022TwistedTemperature, Mishra2020SpinPolymers, Mondal2020Long-RangeMagnetization, Lu2021HighlyHalides}, can be regarded as the expression of stochastic hysteresis between the slow and fast transport channels. The expression of multi-stability of the transported current magnitude is a transient phenomena, guided by a complex interference of localized and delocalized charges. As the timescale of the slow channel has a subtle influence on the fast channel, the system is continuously switching between the two transport channels. An expression of this behaviour must be found in the voltage sweep rate applied to the metal-molecule-metal junction \cite{Migliore2013IrreversibilityJunctions}. The occupation state of the redox center which influences the transmission through the charge channel can hence be probed. 

We ramped the bias at different sweep time per IV curve. A \SI{20}{\nano\meter} thin crystal on HOPG was probed. \textbf{Fig. \ref{fig:C-AFM}G} and \textbf{\ref{fig:C-AFM}H} show \SI{20}{} superimposed IV curves taken at 1.4Hz  and 15.6 Hz, respectively. The former correspond to a slow sweep of the IV curve, while the latter to a fast sweep. Between the two, we observe behaviour that is fundamentally different, although taken at exact same spot and tip force on the molecular crystal. In \textbf{Fig. \ref{fig:C-AFM}G}, we observe no hysteresis and almost smooth monotonic IV curves. Only a few spikes are observed. For this situation the sweep rate is slow enough that the transport current can adiabatically follow the bias potential, and the memory effect induced by the redox site is almost nullified \cite{Migliore2013IrreversibilityJunctions}. Contrary, by sweeping fast (\textbf{Fig. \ref{fig:C-AFM}H}), a significant hysteresis is observed and the spikes are present indicating the redox site localises carriers within the model of Migliore and Nitzan  \cite{Migliore2013IrreversibilityJunctions}. The observed peaks or Fano resonances are equidistantly spaced indicating the resonant behaviour of carrier capture and release. We likely only observe this periodic behaviour \textbf{Fig. \ref{fig:AFM_Crystals}B, C} and \textbf{Fig. \ref{fig:C-AFM}G} for thin crystalline layers, i.e. thinner than \SI{20}{\nano\meter}. For thicker crystals the transport disorder introduced by the different localization rates introduce significant decoupling of the peak spacing and hence more disordered spacing of the peak (\textbf{Fig. \ref{fig:C-AFM}F}). 

\begin{figure*}
    \centering
        \includegraphics[scale=0.48]{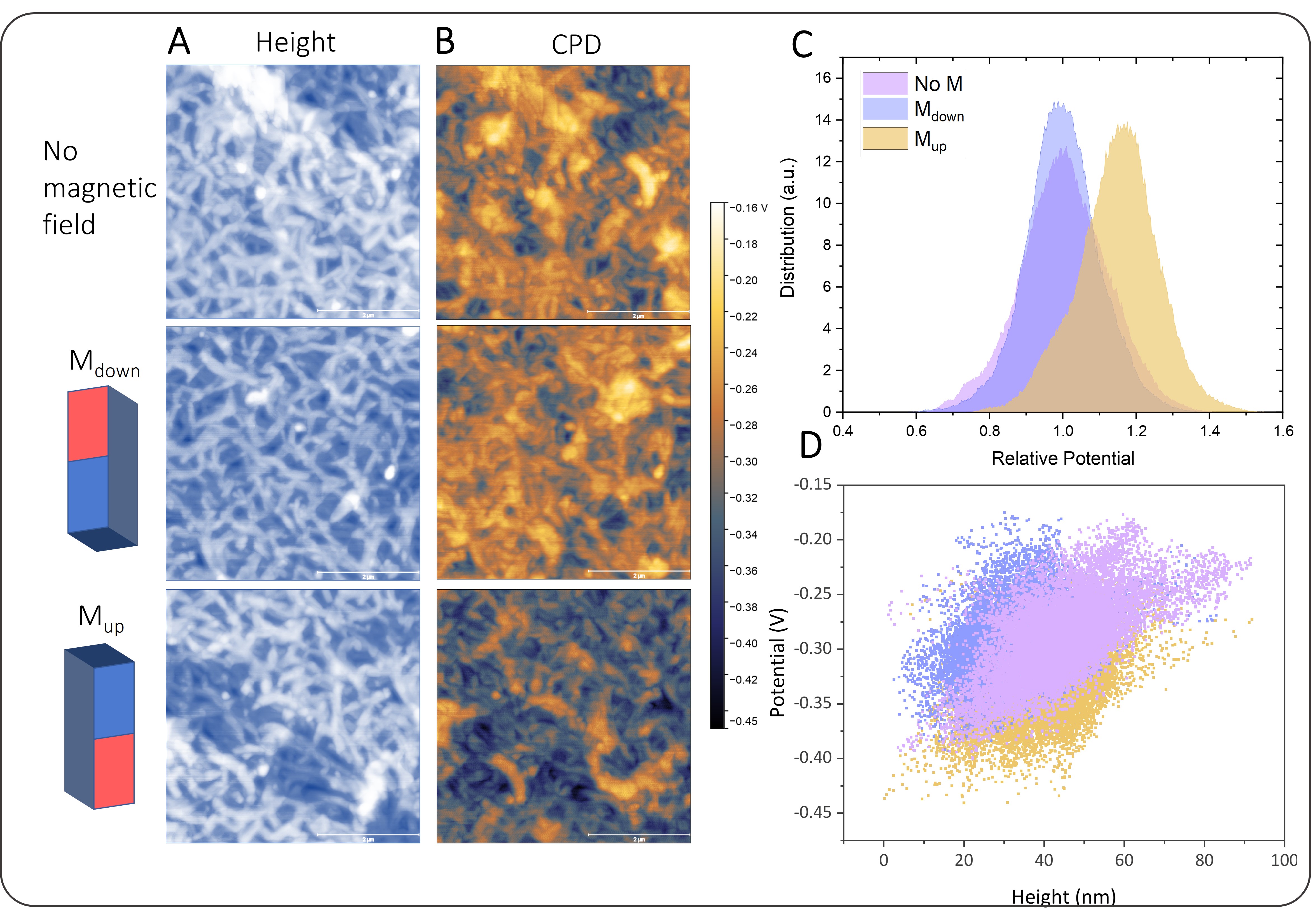}
            \caption{\textbf{Spin filtering by contact potential shift under application of magnetic field}
            (\textbf{A}, \textbf{B}) Topographic, left column and contact potential difference (CPD), right column, measured as function of no magnetic field, and downward/upward magnetic field. A relative shift in CPD is evident for only upward pointing magnetic field. (\textbf{C}) Statistical distribution of the CPD of \textbf{B} as function of magnetic field. (\textbf{D}) Plotting all CPD versus height data of \textbf{B} shows a relatively large spread and approximately linear increase as function of height.}
                \label{fig:MCP}
\end{figure*}

\subsubsection{Probing spin-polarisation of CBI-GCH crystals}
Having established the structure and transport characteristics of CBI-GCH crystals, we turned to probing the spin-polarisation possibility of the crystalline aggregates. We measured the change in contact-potential difference (CPD) with electrostatic force microscopy (EFM) following work from Naaman \textit{et al} \cite{Ghosh2020EffectInterfaces}. The authors successfully demonstrated  observing spin-dependent electronic resistance variation at metal-chiral molecule interfaces by employing (a)chiral self-assembled organic monolayers (SAMs) \cite{Ghosh2020EffectInterfaces}. The benefit of this method is that no tip damage is incurred by the non-contact nature of EFM. A fundamental relation between dipole orientation and chirality induced spin filtering was also found Eckshtain-Levi \textit{et al.} \cite{Eckshtain-Levi2016ColdMonolayers}.  In this work, we performed similar EFM measurement and deposited the CBI-GCH crystals on a gold (\SI{60}{\nano\meter}) coated mica (\textbf{Fig. \ref{fig:MCP}A}). Note, the large SOC of gold is likely a source for SP \cite{Alwan2021SpinterfaceEffect}. The CPD is measured as function of no magnetic field (No M) and the magnetic field point "up" ($\text{M}_{\text{up}}$) and "down" ($\text{M}_{\text{down}}$). Because we have to manually invert the permanent magnet placed below the sample, identifying the exact same topographic area is difficult. However, the experimental data is taken from the  same area of the molecular aggregate within \SI{5}{\micro\meter} positional accuracy. 

The crystals show a significant asymmetric CPD shift of several tens of mV, depending on the relative orientation of the external magnetic field, (\textbf{Fig. \ref{fig:MCP}B}). This confirms that the crystals are spin-polarisers, akin to earlier reports for 1D supramolecular polymers \cite{Kulkarni2020a} and other EFM reports \cite{Ghosh2020EffectInterfaces}. Because the crystals are stacked and thus the molecular aggregate layer thickness varies, an distribution in the magnetic field dependent CPD is observed. Hence,  we analysed the CPD data of \textbf{Fig. \ref{fig:MCP}B} to obtain the statistical distribution (\textbf{Fig. \ref{fig:MCP}C}). The statistical distribution shows a relative shift of \SI{20}{\percent} of peak position for the upward orienting magnetic field, with respect to the no magnetic field. 
We must note that the spread in distribution is relatively large, and overlap of the CPD between all magnetic states exist. This could hint the ordering of the crystals within the aggregates being important for superposition or reduction of the total polarisation. Also, for increasingly thicker films, the CPD shifts approximately linear (\textbf{Fig. \ref{fig:MCP}D}). This indicates an internal polarisation/dipole is observed, which we relate to chiral nature of the molecular aggregate.

\subsubsection{Model}  
\begin{figure*}
    \centering
       \includegraphics[scale=0.5]{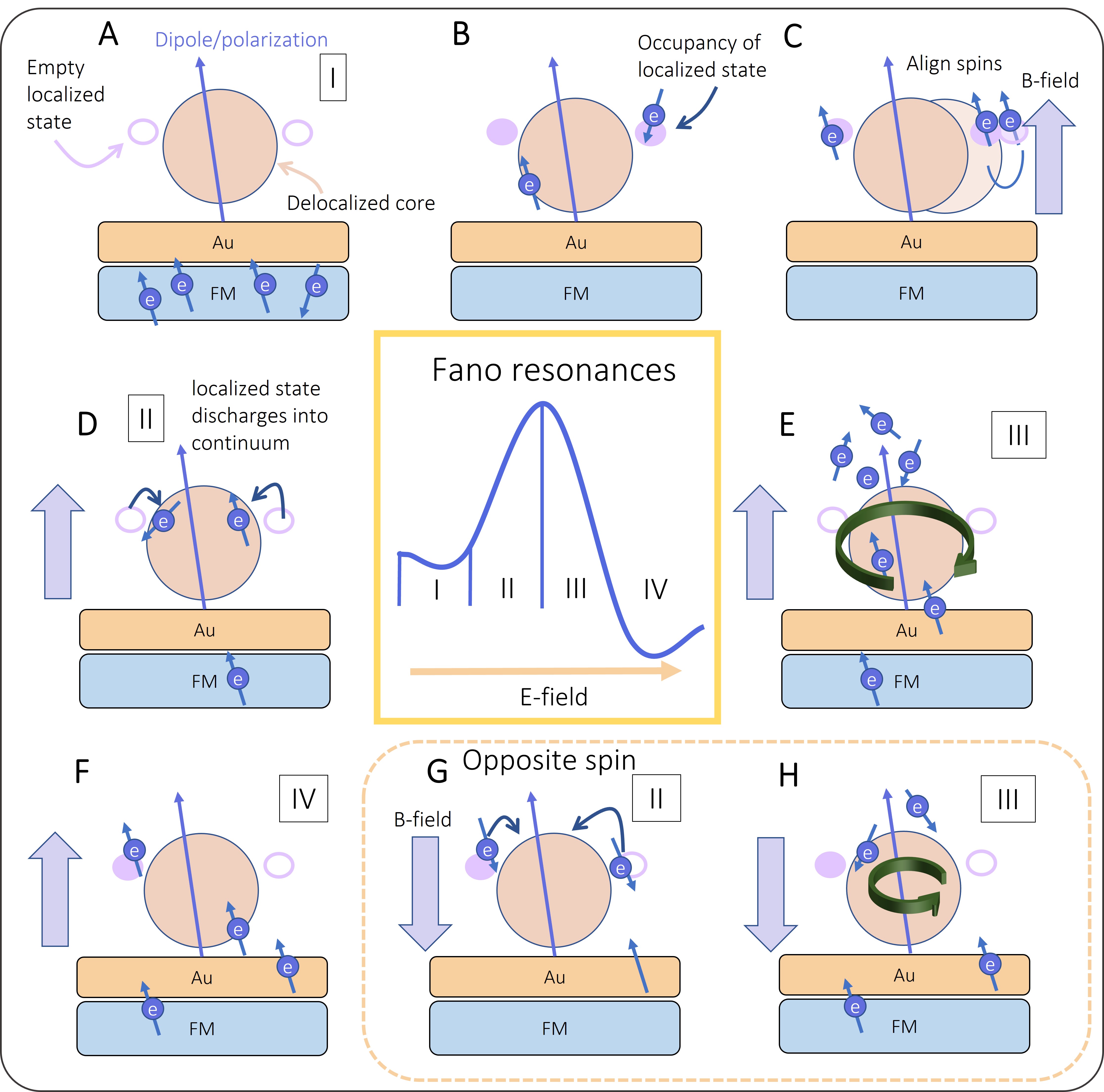}
            \caption{\textbf{Schematic model indicating transport modulation driven by imide redox sites of CBI-GCH.}
            (\textbf{A}) Localized states are close and connected to a continuum. The molecule posses a dipole and the direction and polarisation is dependent on the handedness and adsorption on a metallic substrate. A ferromagnetic (FM) layer may be present to inject spin-polarised charge. (\textbf{B}) Ramping the bias can lead to charge injection and trapping at the redox sites. (\textbf{C}) Spins can couple via exchange interaction with neighbouring trapped charges, on the parallel discotics, and aligned in an external B-field. (\textbf{D}) Further ramping of the bias eventually crosses the Fermi-level into the LUMO, and leads to coupling with the continuum. (\textbf{E}) Charge displacement from the redox sites leads to spin order loss, and an induction current is generated through the continuum. The magnitude of the induction current is related to the sum of the external bias and internal polarisation vector. (\textbf{F}) Further ramping of the bias leads to the downward slope of the Fano resonance and blocking of current flow. (\textbf{G}) Opposite spin localisation. (\textbf{H}) The delocalisation of the localised spins into the continuum lead again to an inductive current, but in reduced magnitude due to the polarisation vector subtracting the total electric field across the molecule.}
                \label{fig:Model}
\end{figure*}

We turn our attention to a novel mechanism to correlate CBI-GCH origins of charge dynamics \cite{Migliore2013IrreversibilityJunctions} and the reported high degree of spin polarisation of CBI-GCH \cite{Kulkarni2020a}. Migliore and Nitzan \cite{Migliore2013IrreversibilityJunctions} describe the charge transfer phenomena within transition rate process models. Here we suggest a novel mechanism for a microscopic understanding of the measured charge dynamic effects (\textbf{Fig. \ref{fig:Model}}).  Indicative expression of the model are what we assign as Fano like \cite{Nozaki2013AJunctions} peak structures. A localised state and a continuum state, via the delocalised core of the molecule (\textbf{Fig. \ref{fig:Model}A}) govern the charge transfer. The continuum could be represented by $\pi$ - $\pi$ hybridization of the stacked coronene cores. Another possibility is the novel mechanisms of highly diffuse molecular orbitals (HDMO). We note that for coronene, HDMO orbitals have been predicted and experimental verified \cite{Ortiz2020Dyson-orbitalMolecules} as they can arise in the excited state \cite{Pastukhova2017EvidenceFilms}. The HDMO share similarities with the super-atomic molecular orbitals (SAMO) observed for curved aromatic molecules \cite{Pastukhova2017EvidenceFilms} and fullerenes \cite{Feng2008Atomlike60}. In our case the HDMO requires a dipole, which is provided by the chirality induced by the helix and screening from the metallic substrate. A ferromagnetic (FM) layer may be used to inject SP carriers, but is not a requirement \textit{per se}. 

For CBI-GCH in the ground-state, the LUMO-0 is located near the redox imide (\textbf{Fig. \ref{fig:AFM_Crystals}A}). The redox sites can be half-filled as on-site Coulomb repulsion between the stacked redox sites of the stacked discotics, will prevent double occupation and form electron states very close to the LUMO level. For the thin crystal structures embedded between two metal contacts, a fast increase of the bias voltage (\textbf{Fig. \ref{fig:Model}B}, \textit{State I}) above the bandgap will fill the localised redox states. The closeness of the redox sites (\textbf{Fig. \ref{fig:AFM_Crystals}A}) can align the spins in a particular direction due to exchange interaction. In addition, a magnetic field can also be applied to align to spins of the localised charge (\textbf{Fig. \ref{fig:Model}C}). Further increase of the bias voltage will allow the continuum like states to be filled and coupled to the localised states to form a Fano resonance \cite{Wang2021SpinImplications} (\textbf{Fig. \ref{fig:Model}D}, \textit{State II}). The Fano like resonance can also be described by the Anderson impurity model \cite{Anderson1966TheoryMetals}. Within this model the probability of double occupancy of the localized state is strongly reduced due to the on-site Coulomb repulsion and the other term in the Hamiltonian represents the hybridization between the redox pendant state and the continuum. 

The continuum will boost the current density so much that a significant spin-orbit coupling may be induced, which in the presence of the redox sites, can introduce a spin-dependent current through the stack of the CBI-GCH molecules for a short time before the localised charge will become part of the continuum (\textbf{Fig. \ref{fig:Model}E}, \textit{State III}) \cite{VanBree2014Spin-orbit-inducedNanostructure, RodriguesdaCruz2022DissipationlessGas}. This would implicitly mean that the metal electrodes in the transport set-up need not necessary to be magnetic for observing the CISS effect in CBI-GCH molecular systems, supporting the observed CPD shift on gold/mica substrates. 
Hereby, the classical spin-orbit coupling state of the continuum current contribution with an additional part from the orbital current, followed by a current dip induced by the spin-order loss (\textbf{Fig. \ref{fig:Model}F}, \textit{State IV}), mimics the Fano shape behaviour. The low conductivity is finally restored and new charges can be trapped at the redox sites (\textbf{Fig. \ref{fig:Model}F}, \textit{State IV}). For opposite spin, similar mechanism occurs (\textbf{Fig. \ref{fig:Model}G}, \textit{State II}) with localisation on the redox groups. However, the inductive current magnitude into the continuum (\textbf{Fig. \ref{fig:Model}H}, \textit{State IV}) is lower due to the polarisation vector leading to a subtractive current density. The interpretation of charge transport in this way provides a spin-induced contribution to the total current, depending on the spin orientation in relation to the main current direction, hence supporting the spin polarised induced current density difference in the way as has been measured in the CISS experiment of \cite{Kulkarni2020a}. In \textbf{SuppS10} the spin polarisation model of enantiomers is discussed. 

\section{Conclusions $\&$ Outlook}

The success attained here to unravel the reason for the spiky IV spectra measured with a conductive AFM which opens a way to understand the fascinating CISS effect. The clue is the combination of continuum, perhaps a HDMO state, and a strongly localized state by a pendant redox group, which provides the asymmetry in the conduction of spin polarized charge. Similar systems have been created by the coupling of paramagentic molecules to carbon nanotubes yielding spin polarisation \cite{Hong2013Fano-resonance-drivenMagnets}. Localisation of charge and the coupling between them generates a spin-polarised molecular state. Electronic coupling to a continuum allows for delocalisation by a Fano resonance.  We argue spin polarization is induced by delocalisation of the trapped charge on the redox site into the continuum, generating a large “inductive” current, depending asymmetrically on the polarisation vector driven by the molecular handedness. Future work of electrochemical gating may be uesd to actively control the redox state of the diimide \cite{Li2015Three-StateTuning} and thus possibly control the degree of spin-polarisation.  

This new view on the role of chirality and spin polarization offers novel insights to define design rules for chirality supramolecular systems and applications for magneto resistance with a high degree of spin polarisation at room temperature. A strong continuum may be created by actively introducing a super atomic molecular orbital (SAMO) close to the LUMO level \cite{Pastukhova2017EvidenceFilms, Bohl2017Low-lyingAlkanes, Feng2008Atomlike60}. Such orbitals benefit from the highly delocalised orbital nature beyond the dimension of the molecule. This way, free-electron like conductivity can be introduced for slightly curved aromatic molecules when adsorbed on metal surfaces. By including charge localisation states close the continuum, i.e. by redox pendent sites, an ideal spin-polarised Fano system may be created based on chiral molecules.

\section{Materials and Methods}

\subsection{Substrates}
Synthesis of CBI-GCH is described in \cite{Kulkarni2013Self-assemblyAmplification}. Freshly cleaved HOPG substrates where used for the deposition of the crystals for IV spectroscopy. For EFM, the crystals where drop-casted on gold (\SI{60}{\nano\meter}) evaporated mica sheets. The mica sheets where prior for coating freshly cleaved. Drop-casting was performed by elevating the sample temperature to \SI{40}{\celsius} to increase the evaporation rate of MCH and thus enhance the local crystalline concentration. Several drop-cast were made consecutively. ITO-coated glass substrates were evaporated with \SI{10}{\nano\meter} of gold and CBI-GCH drop-casted to with the sample held at \SI{40}{\celsius}. 

\subsection{CD measurements}

Circular Dichroism (CD) measurements were performed using JASCO J-815 CD spectrometer using the following settings; sensitivity: Standard, D.I.T: \SI{0.5}{\second}, bandwidth: \SI{1}{\nano\meter}, scanning speed: \SI{100}{\nano\meter\per\minute}, data pitch: \SI{1}{\nano\meter}. Bulk measurements were performed on \SI{1}{\centi\meter} x \SI{1}{\centi\meter} x \SI{0.1}{\centi\meter} (l x w x t) quartz slides covered with ITO and \SI{10}{\nano\meter} of gold. Solid state measurements were performed using a solid-state holder with a circular opening of \SI{12}{\square\milli\meter} (d=\SI{4}{\milli\meter}). The presented CD spectra are the average of two spectra measured from front and back to limit the contribution of linear dichroism and linear birefringence to the emergent CD.

\subsection{TEM}
For TEM sample preparation a 200-mesh copper grid covered with a Quantifoil R 2/2 holey carbon film and a \SI{2}{\nano\meter} continuous carbon layer (Quantifoil Micro Tools, GmbH) was surface plasma treated for \SI{5}{\second} using a Cressington 208 carbon coater. Subsequently, \SI{20}{\micro\liter} of the aqueous sample was drop casted onto the carbon film of the copper grid, followed by drying in air. CryoTEM imaging was carried out on the cryoTITAN (Thermo Fisher, previously FEI), equipped with a field emission gun (FEG) and a 2k $\times$ 2k Gatan US1000 CCD camera. The microscope was operated at 300 kV acceleration voltage in bright-field TEM mode at a nominal magnification of \SI{5400}{} $\times$ with a \SI{1}{\second} image acquisition time. Selected Area Electron Diffraction (SAED) pattern were acquired by operating the TEM at LN2 temperature, in diffraction mode using a camera length of \SI{1.15}{\meter}, a selected area aperture of \SI{40}{\micro\meter} and a \SI{0.25}{\second} exposure time.

\subsection{Conduction-AFM}
C-AFM was performed with a diamond tip (Apex Sharp) from Adama Innovations, with a nominal tip radius of \SI{10}{\nano\meter} and a spring constant of \SI{2.6}{\newton\per\meter}. Identification of drop-casted clusters was performed with the AFM optical microscope. The cantilever was gently brought into the soft contact with the molecular cluster, and then retracted to break tip-sample contact. We define soft contact as the point of snap-into-contact of the force-distance curve. The voltage of the piezo tube was continuously recorded and used for force conversion. The tip was placed in soft contact mode to scan the local morphology. This ensured a local pristine morphology to perform IV spectroscopy. Afterward, a larger morphology was imaged to identify the local area probed during spectroscopy. At each force set-point the tip was allowed to stabilize. Little drift was observed as imaging the area after IV spectroscopy highlighted the same molecular topographic features. Bias was applied to the sample. Up to \SI{10}{\pico\ampere} DC off-set was registered as leakage current from the op-amp and subtracted from the IV spectroscopy. The number of data-points and bias sweep rate were registered for each IV curve. 

\subsection{PeakForce AFM}
A Bruker Edge AFM was used with PeakForce Tapping feedback. AFM silicon nitride PeakForce-HIRS-F-B tips (Bruker) with a spring constant of \SI{0.12}{\newton\per\meter} and a nominal tip radius of \SI{2}{\nano\meter} were used. Scan parameters were optimized using Bruker ScanAsyst\textsuperscript{TM}. Before each measurement the drift was nullified by scanning for an hour before measuring the high resolution imaging at the nanoscale. All data were plane fitted to subtract the background offset using Gwyddion Software (http://gwyddion.net). 

\subsection{Electrostatic force microscopy}

A Dimenson Veeco Multimode running a NanoScope III  controler was used for EFM. Pt coated Si tips where used with a nominal tip radius of around \SI{20}{\nano\meter}. The alternating bias voltage was applied to the tip of \SI{6}{\volt} at the eigenfrequency of the cantilever. The tip was retracted by \SI{5}{\nano\meter} to measure the CPD in lift mode. 
In the work of Ghosh \textit{et al.} \cite{Ghosh2020EffectInterfaces} the measurement was performed without actively grounding the metallic film. We grounded the substrate via the mica bottom, similar to the approach in reference \cite{Ghosh2020EffectInterfaces}. The magnetic field was applied via an external permanent magnet placed directly below the sample ($B = $\SI{500}{\milli\tesla}) as measured with a Gauss Probe. All data were plane fitted to subtract the background offset using Gwyddion Software (http://gwyddion.net) 

\subsection{DFT calculations}

Density functional theory simulations were performed using VASP. The PBE exchange-correlation functional was used in conjunction with the projector augmented wave approach. All structures were optimized to their local minima using the conjugate gradient algorithm as implemented in VASP. The cut-off energy for the plane-wave basis set is 400 eV. The Brillouin zone was sampled using a 1x1x1 Monkhorst-Pack grid (Gamma-point only). Electron smearing was employed using Methfessel-Paxton smearing with a smearing width ($\sigma$) of 0.0005 eV. To avoid the spurious interaction of neighboring super cells, the unit cell embedding the oligomers was constructed such that a vacuum layer of at least 10 Å is present in each cartesian direction. It was explicitly verified that the electron density and potential approach zero at the edges of the unit cell
\section{Acknowledgement}

We thank Omur Gokniçar for assistance with the AFM experiments and Riccardo Ollearo for fabrication of the ITO/Au substrates. We thank Mark de Jong for support with the Bruker PeakForce AFM. We thank Chidambar Kulkarni for the synthesis of CBI-GCH supramolecular polymers. 

\section{Supplementary}

\subsection{S1 - AFM image of micrometer scale crystal}

\begin{figure*}
    \centering
        \includegraphics[scale=0.5]{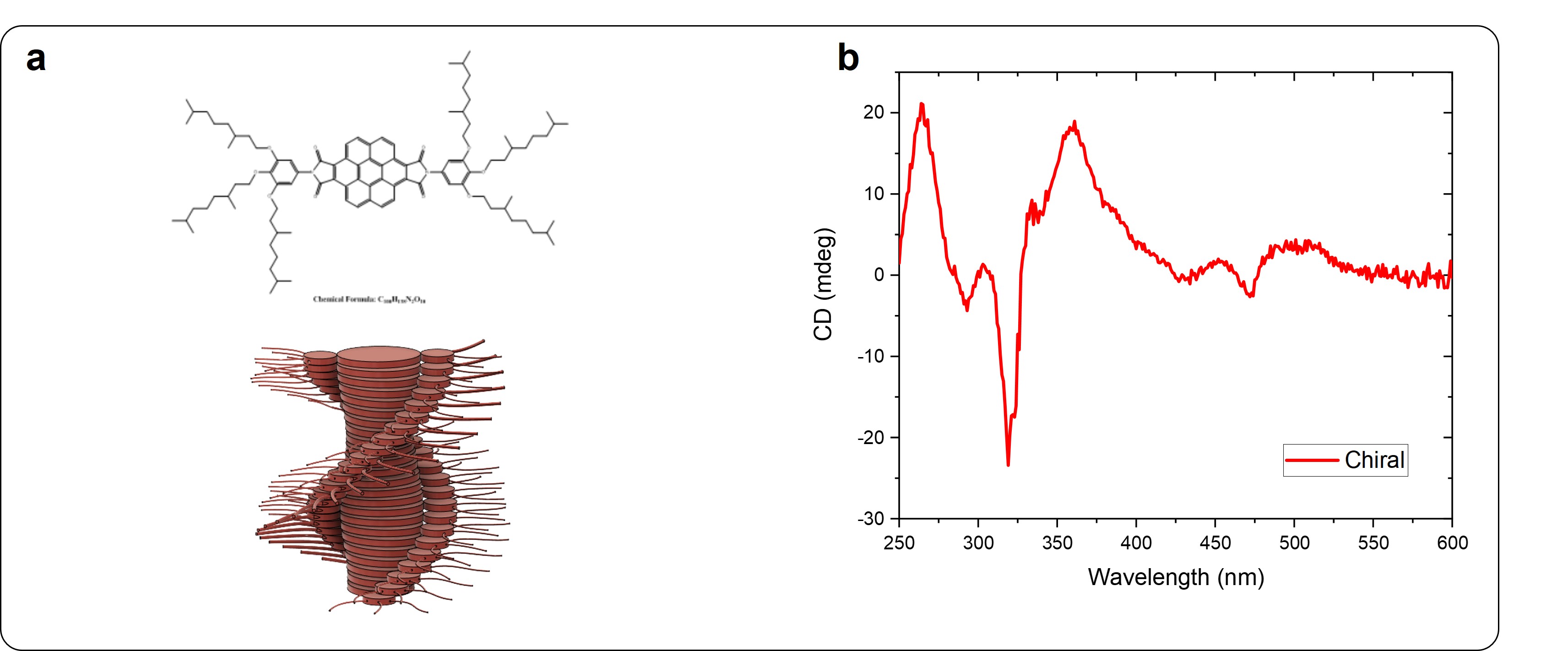}
            \caption{\textbf{Structure and CD of CBI-GCH} 
            }
\end{figure*}

\subsection{S2 - Crystal clustering}

\begin{figure*}
    \centering
        \includegraphics[scale=0.6]{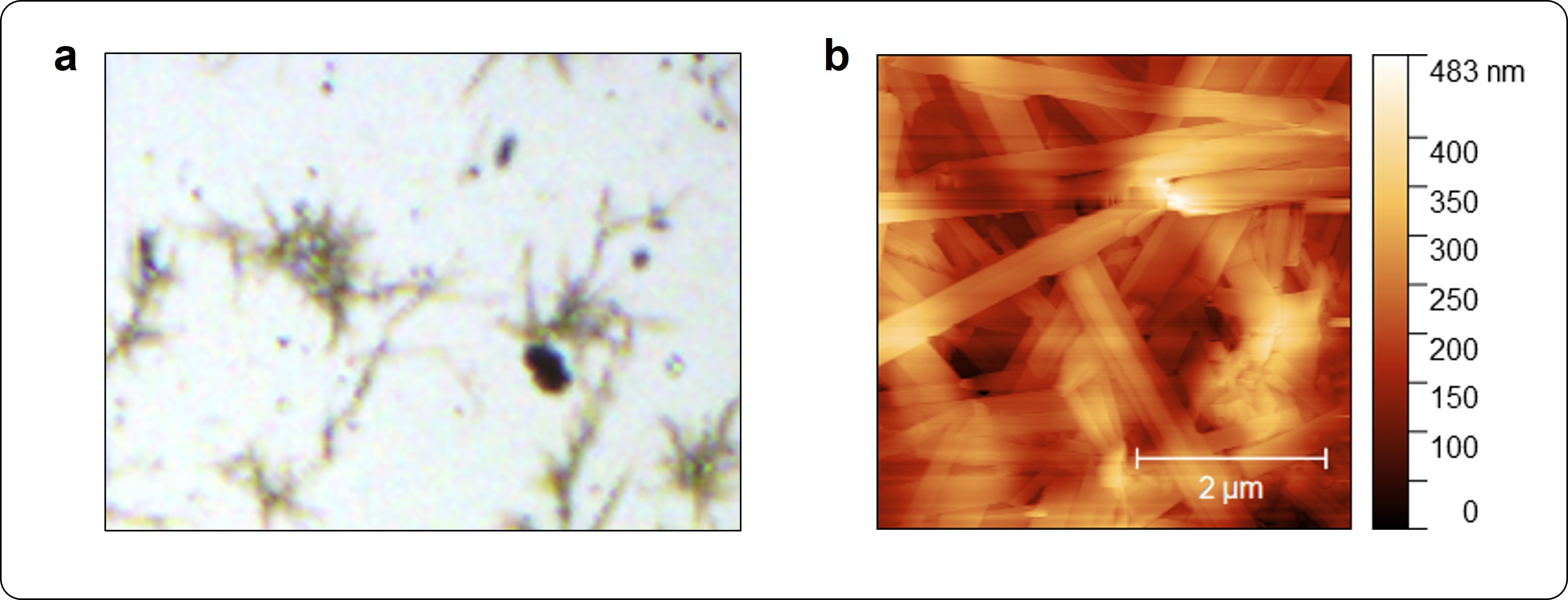}
            \caption{\textbf{Optical and AFM image of clustered CBI-GCH crystals drop-casted on Sio2/Si wafer} 
            }
\end{figure*}

\newpage

\subsection{S3 - AFM image of micrometer scale crystal}

\begin{figure*}
    \centering
        \includegraphics[scale=0.45]{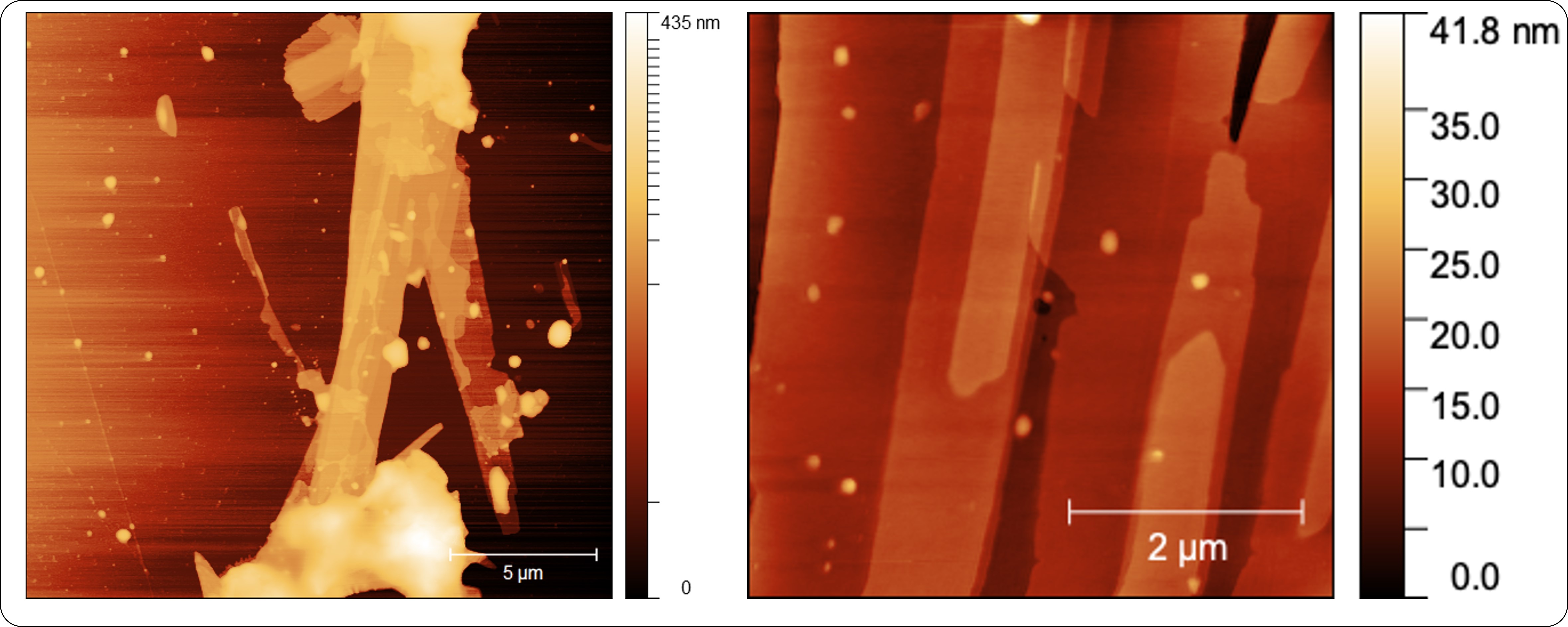}
            \caption{\textbf{AFM topography of micrometer scale crystals deposited on SiO2/Si wafer} 
            }
\end{figure*}

\subsection{S4 - PeakForce spectroscopy}
Force distance spectroscopy was taken on a CBI-GCH crystal deposited on SiO2/Si wafer. The deflection curve shows the snap-to-contact induced by the attractive forces (light green). When the tip is retracted, a significant adhesion force is noted across more than \SI{100}{\nano\meter} of tip retraction. Clearly, a component of the CBI-GCH crystals are sticking to the tip, preventing snap-from-contact. Only at very large distances, in excess of \SI{120}{\nano\meter} the pulling force is released and the tip oscillates (ringing). We interpret this behaviour from alkyl chains sticking to the AFM tip after contact and adhering when retracting.

\begin{figure*}
    \centering
        \includegraphics[scale=0.55]{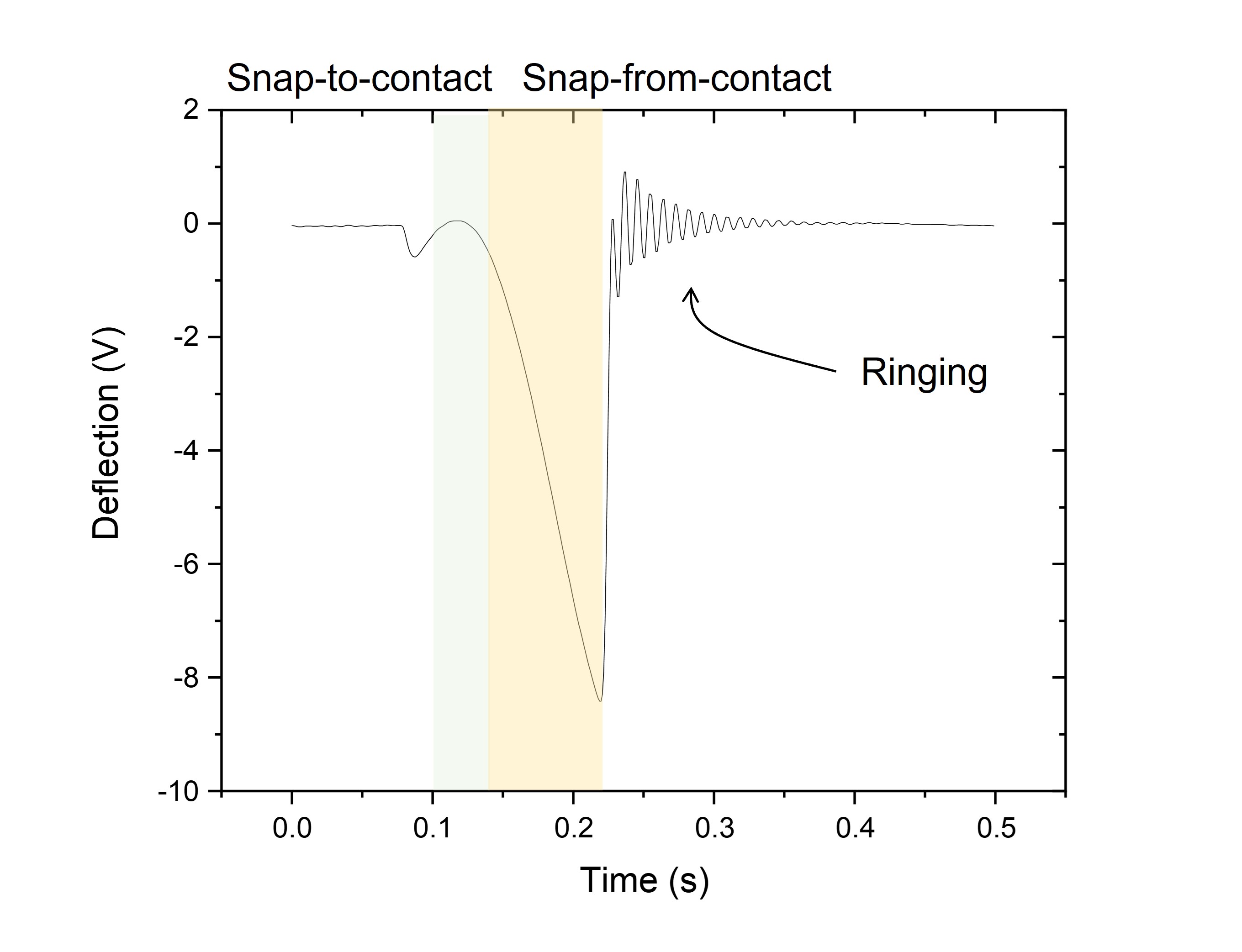}
            \caption{\textbf{PeakForce AFM force-distance spectroscopy} 
            }
\end{figure*}

\newpage
\subsection{S5 - TEM}

\begin{figure*}
    \centering
        \includegraphics[scale=0.42]{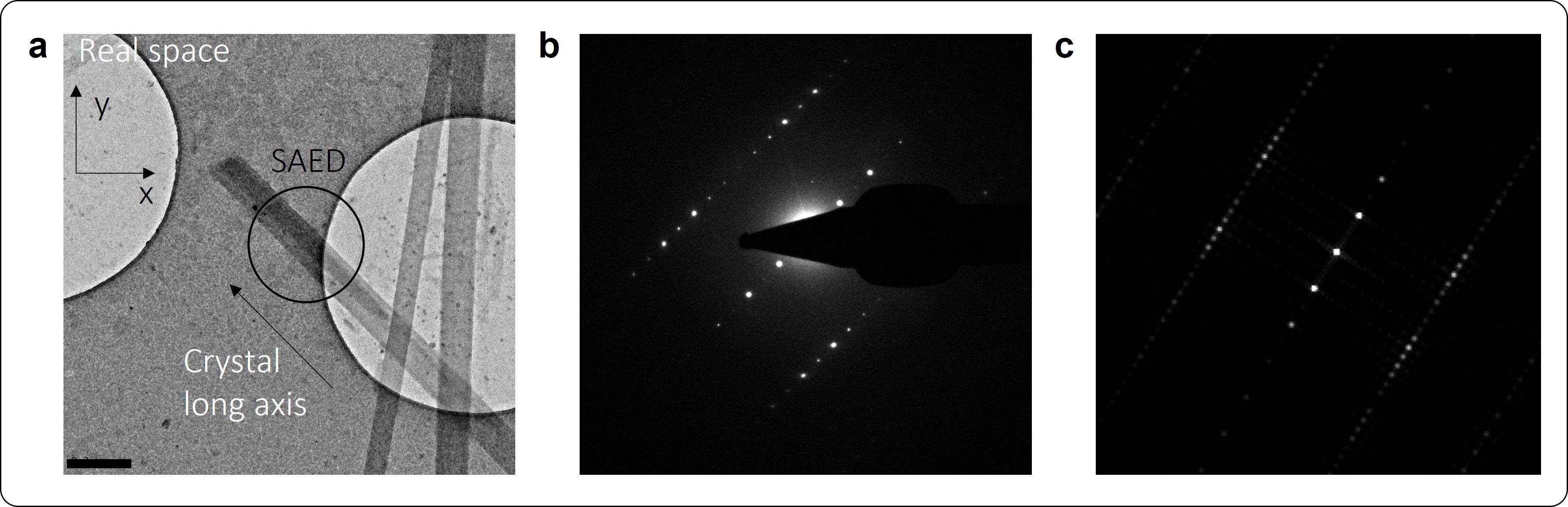}
            \caption{\textbf{TEM imaging}
            (\textbf{a}) Real space image of a crystal deposited on laceycarbon TEM grid. The black circle indicates selected area diffraction (SAED) location. (\textbf{b}) SAED showing high quality diffraction pattern. (\textbf{c}) Simulated FFT of a herringbone structure}
\end{figure*}

\subsection{S6 - Damage to aggregates}

\begin{figure*}
    \centering
        \includegraphics[scale=0.6]{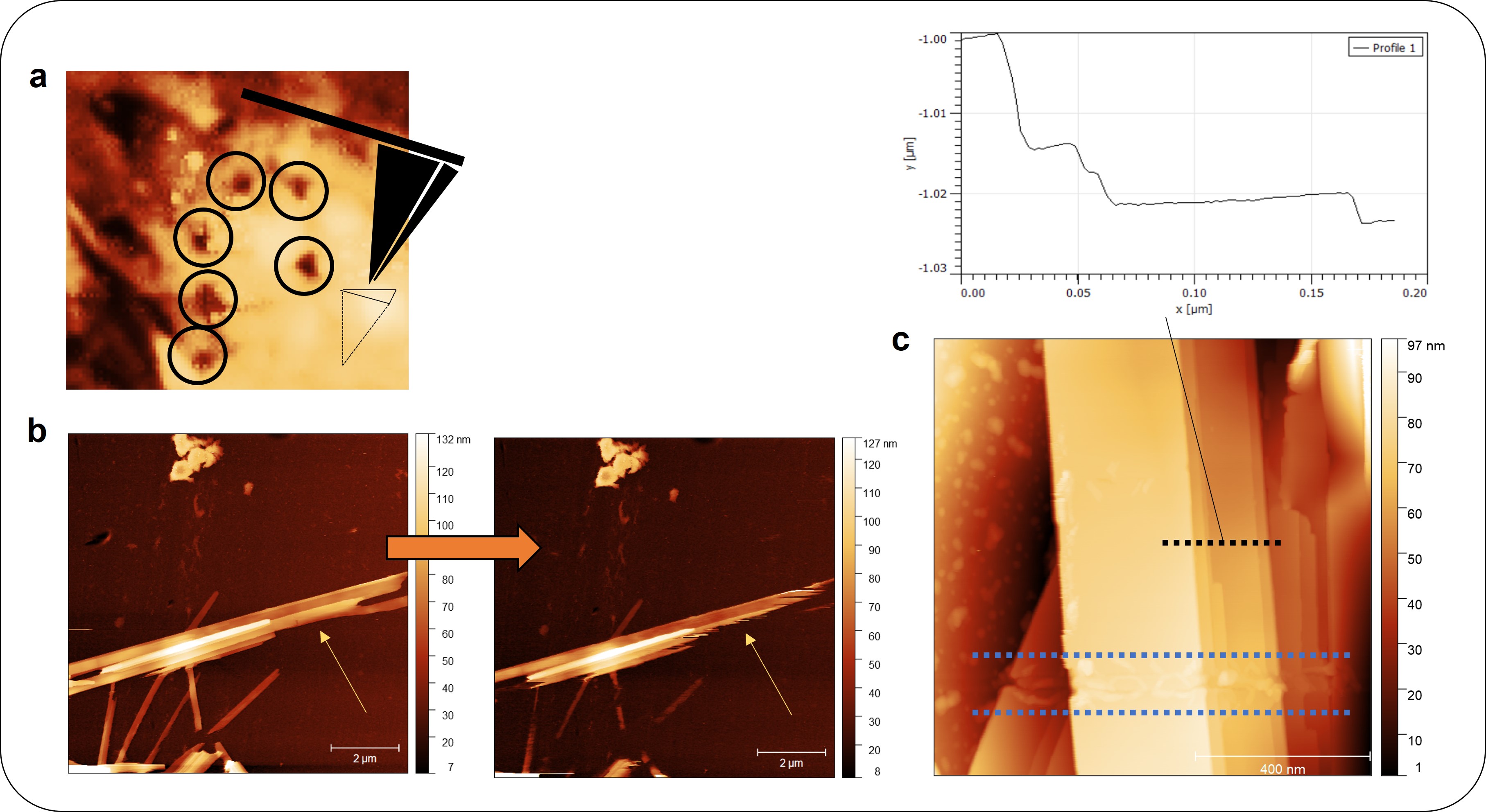}
            \caption{\textbf{AFM induced damage to CBI-GCH aggregates.} 
            (\textbf{a}) Normal force induced tip indentation. (\textbf{b}) PeakForce induced damage to the crystal evident by sequential imaging of the same area. (\textbf{c}) CBI-GCH crystal showing local damage incurred by Amplitude Modulation Tapping Mode between the blue lines, by gently increasing the normal force. The black lines shows a cross-section line highlighting the stacked nature of the crystals.}
                \label{fig:Damage}
\end{figure*}

We observed with tapping AFM, contact AFM and even PeakForce AFM damage was easily incurred onto the supramolecular structures. Figure \ref{fig:Damage} shows topographic images of such damage. Figure \ref{fig:Damage}a shows indentation area's of the diamond tip, gently pressed into the molecular cluster with increasing normal force. The triangular shape of the tip can be observed. Figure \ref{fig:Damage}b shows PeakForce tapping images. We used a Bruker Edge with ultra sharp tips with a radius of \SI{2}{\nano\meter}. The peakforce reduces lateral forces by retracting the tip at each pixel before moving down towards the next pixel during imaging. The yellow arrow shows the complete disappearance of parts of the crystal, simply by imaging the local area again.  Hence, the extremely soft nature of these materials makes application of scanning probe microscopy notoriously difficult. Figure \ref{fig:Damage}c shows an Amplitude Modulation Tapping Mode image. By gently increasing the normal force, the tip was able to cut through several layers of the crystal, highlighted with the blue lines. The stacked nature is evident from the cross-section taken at the black line. 

\subsection{S7 - F-z calibration of C-AFM tip force}

\begin{figure*}
    \centering
        \includegraphics[scale=0.6]{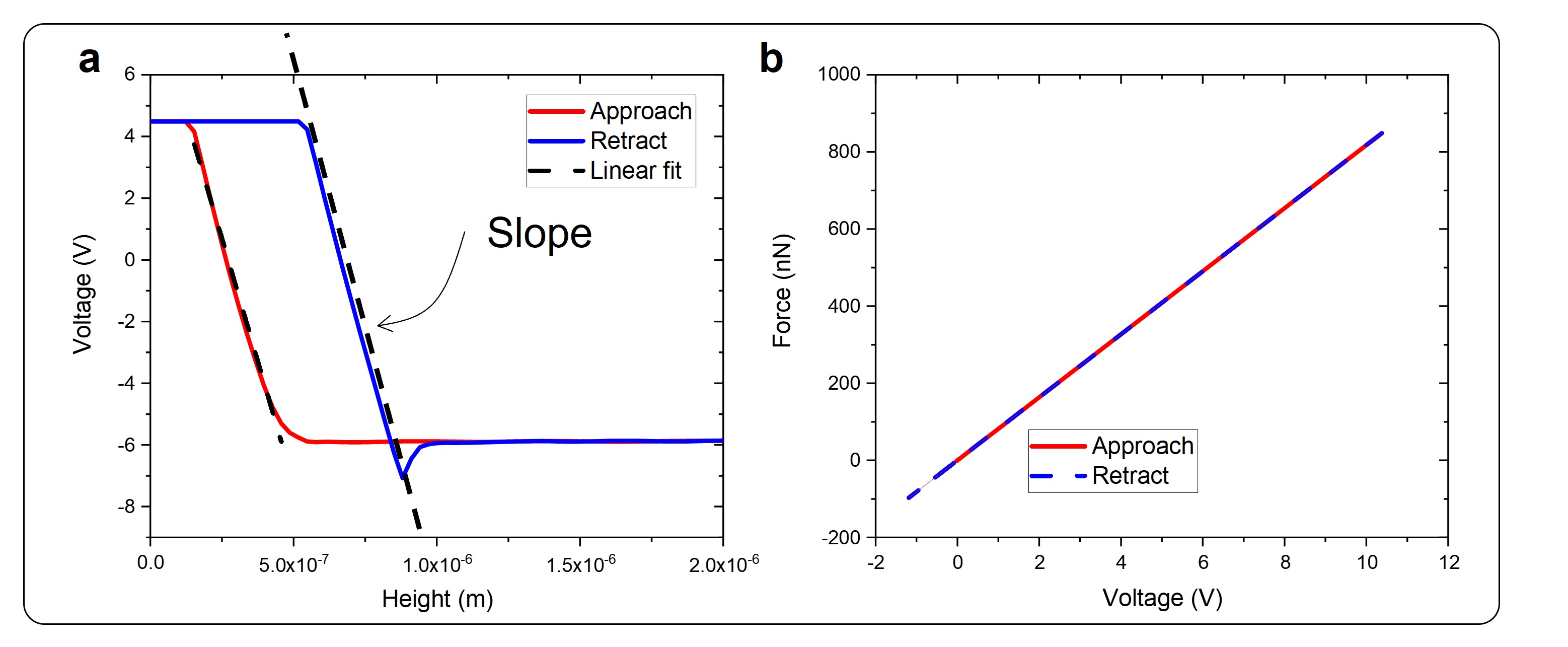}
            \caption{\textbf{F-z cure conversion of force.} 
            (\textbf{a}) F-z curve with approach and retract curve. (\textbf{b}) Converted F-z force to derive the tip force in nN.} 
                \label{fig:Conversion}
\end{figure*}

The force applied by the C-AFM was determined following standard procedure \cite{Sader1995MethodCantilevers}. A F-z curve was taken by placing the tip on the surface, with no molecules present. The tip was approached and retracted and the voltage of the tip deflection registered, see Figure \ref{fig:Conversion}. The voltage-offset was nullified. The force constant was taken from manufacturer specification (AdamaProbes). Figure \ref{fig:Conversion}b shows the linear relationship between the applied tip force and registered voltage. 

\newpage

\subsection{S8 - C-AFM tip state evaluation }

Metal coated Si tips, such as those coated with PtIr, often show continuous degradation between 10's of IV spectra due to mechanical removal or deformation of the metallic film \cite{Jiang2019UnderstandingMicroscopy}.  At bias voltages exceeding several volts, the electric field can easily reach as high as \SI{e8}{\volt\per\meter}, which can introduce non-reversible tip state alteration and thus changing the contact resistance continuously. It is hence important to compare the conductivity of the tip before the experiment and especially after, to exclude artefacts. To reduce the effect of tip degradation, doped single crystal diamond tips (Adama Innovations) have been used throughout this work, which show no degradation even after prolonged AFM scanning \cite{Celano2019TheNanoelectronics}. Furthermore, they enable Ohmic contact and withstand high bias, up to 10V in the used setup, without notable degradation.  

\begin{figure*}
    \centering
        \includegraphics[scale=0.8]{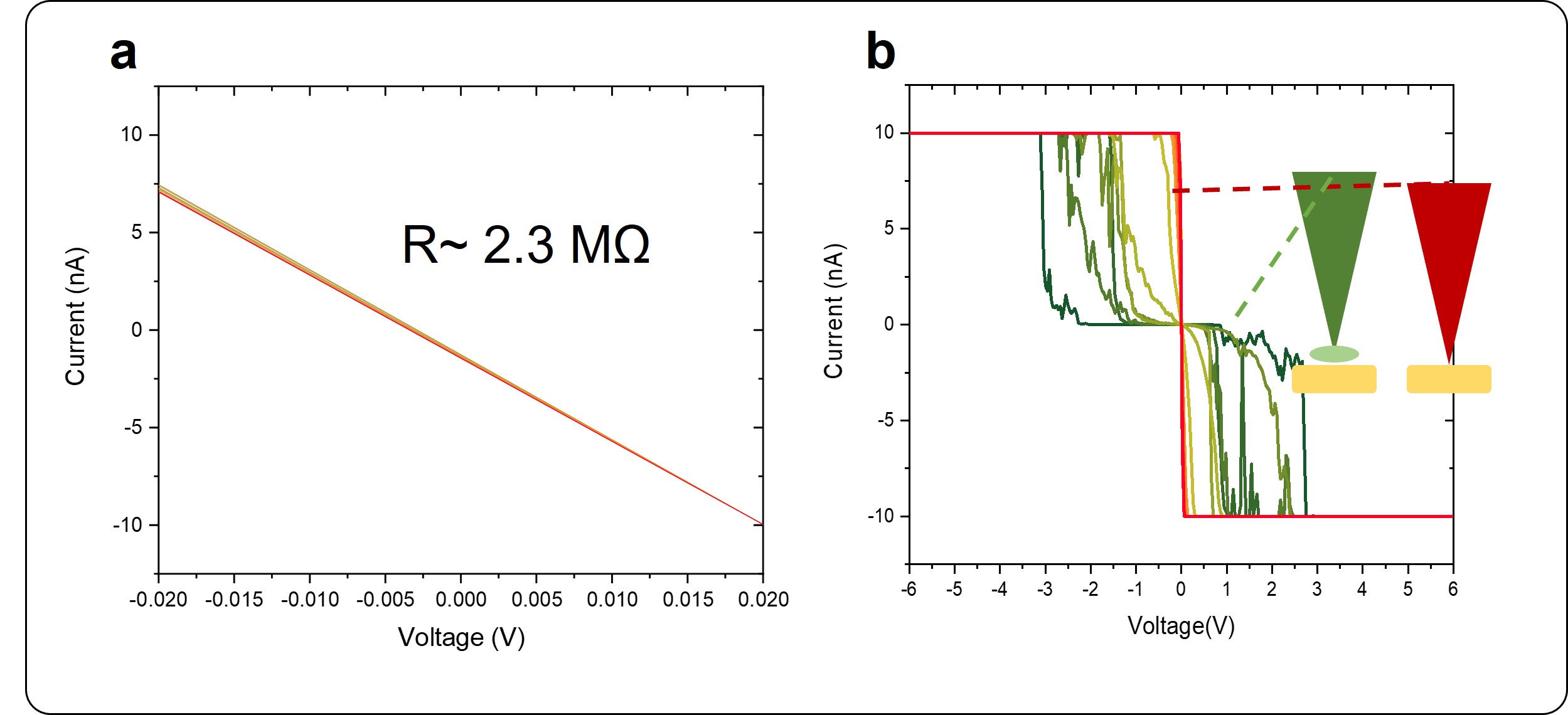}
            \caption{\textbf{IV spectroscopy of molecules coated tip on gold substrate.} 
            (\textbf{a}) Ohmic contact resistance diamond tip on gold. (\textbf{b}) Observation of gradual change in IV behaviour (green) by sequential bias voltage ramping, until Ohmic resistance (red) is achieved. We conjecture tip contamination with molecular clusters.}
                \label{fig:Tip}
\end{figure*}

We observe an intrinsic conductive state of the Adama Innovations diamond tip with Ohmic contact resistance of around \SI{2.3}{\mega\ohm}, see Figure \ref{fig:Tip}a, measured by pressing the diamond tip directly on gold. The IV curves are swept 20 times, with no noticble spikes or anomalies in the spectroscopy. Furthermore, we observe no change in tip conductivity during experiments. We note a small current offset below \SI{100}{\pico\ampere}, because of leakage current from the current-to-voltage amplifier. 

However, adsorption of molecules on the tip is difficult to rule-out. For example, it was observed that after repeating several IV spectroscopy sequences on the molecular bundles and subsequently placing the tip on the gold surface, no direct Ohmic behavior could be established. A clear example of this is given in Figure \ref{fig:Tip}b, the first IV curves, green, shows non-linear IV behavior, even though the tip is directly contacted on gold with no large molecular aggregates present. It could be argued that previous taken IV data on molecular aggregates has resulted in the transfer of molecules to the tip which act as tunneling barrier when contacting gold. Subsequent sequences of spectroscopy shows gradual degradation in the tunnel gap (lighter green) hinting at likewise sequential desorption of molecules likely due to very large electric fields of $10^8 - 10^9$ V/m. A sudden change to Ohmic behavior was established after several IV spectroscopy cycles. This behavior shows that the intrinsic tip state of Ohmic conductivity was indeed unaltered, but the ill-controllable adhesion of molecules is problematic and we conjecture it can be a source of wrongly reporting absolute SP degrees in literature. 

\subsection{S9 - Resonant peaks}
 
\begin{figure*}
    \centering
        \includegraphics[scale=0.7]{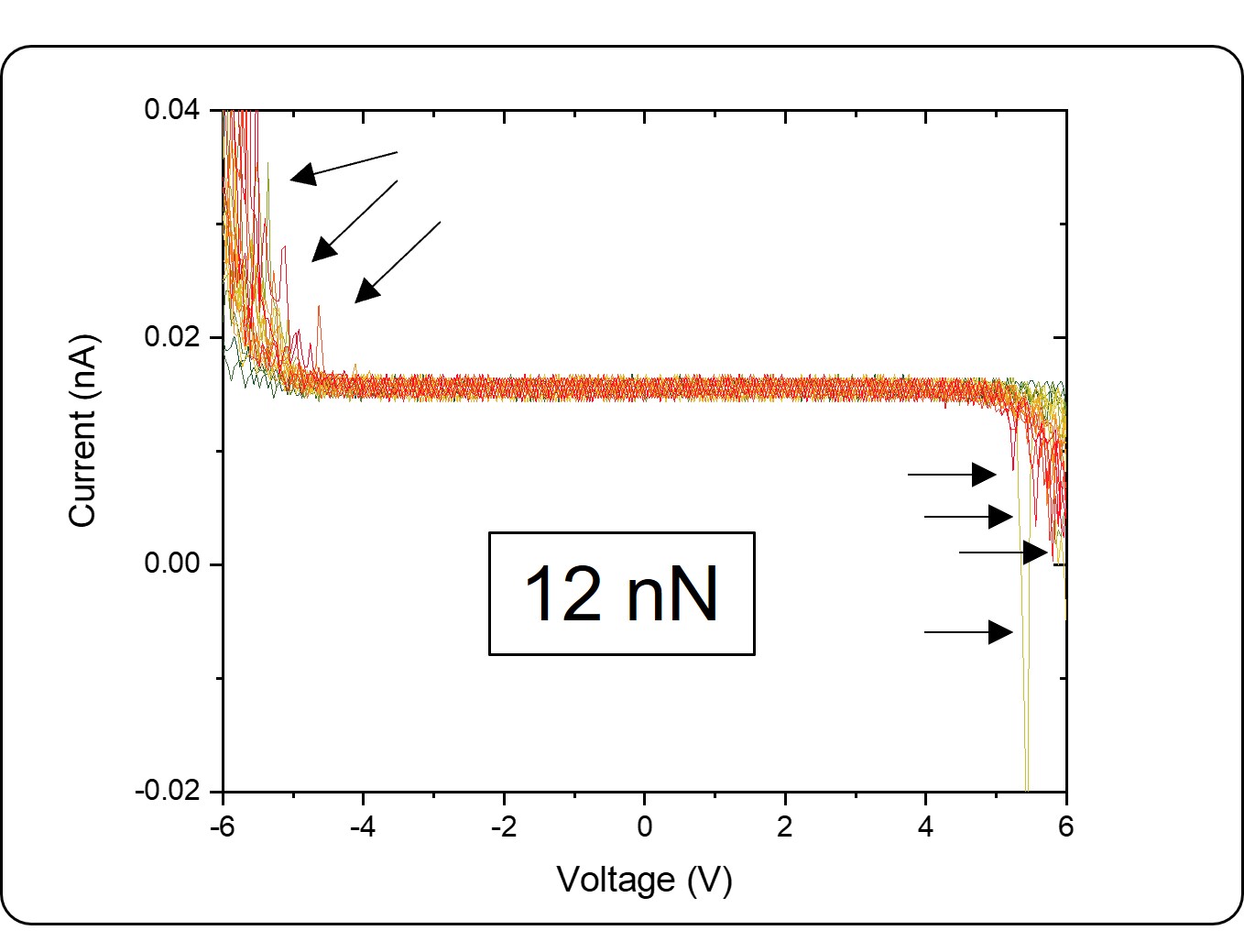}
            \caption{\textbf{IV spectroscopy of a \SI{35}{\nano\meter} thick CBI-GCH crystal.} 
            Black arrows highlight peaks. The tip force is set to \SI{12}{\nano\newton}.}
                \label{fig:Peaks}
\end{figure*}

\subsection{S10 - Spin-polarisation model}
\begin{figure*}
    \centering
       \includegraphics[scale=0.47]{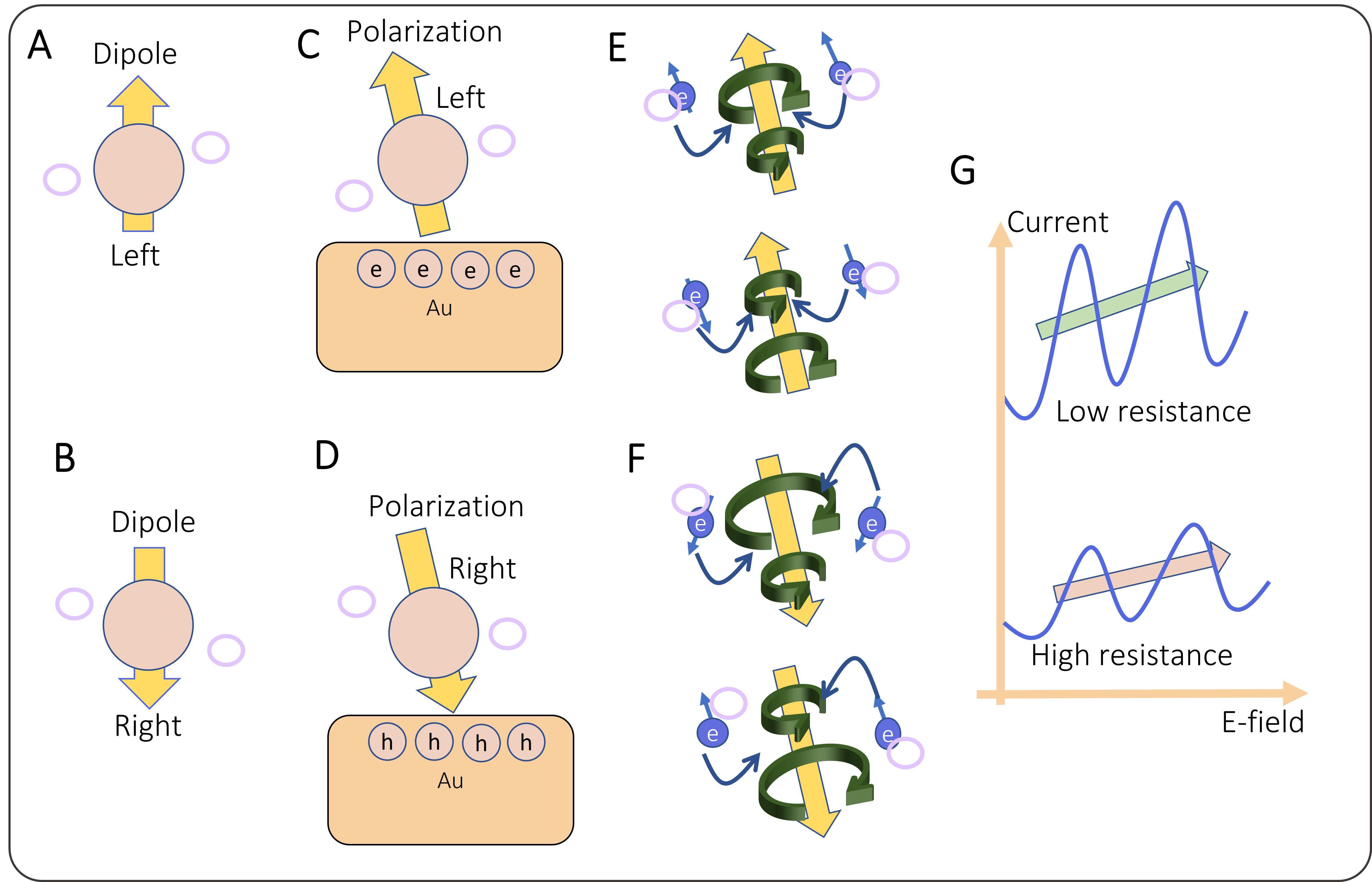}
            \caption{\textbf{Schematic model of the relation between chirality (dipole) and preferential spin-polarisation.}
            (\textbf{A}, \textbf{B}) Chiral molecule with opposite handeness determines relative orientation of dipole. (\textbf{C}, \textbf{D}) Adsorption of molecule on metallic substrate introduces screening and formation of polarisation vector, depending on molecule handedness. (\textbf{E}, \textbf{F}) Charge localisation on the redox, with aligned spin are coupled to the continuum. Depending on the localised spin orientation with respect to the polarisation, on trapped charge delocalisation, an induction current is generated. The inductive current is either additive or subtractive driven by the polarisation vector. (\textbf{A}) The sum of inductive current on the rise and fall slope of the Fano resonance, determines the slope of the current rise. A different slope is expected for preferential spin induced inductive current from the localised state coupling to the continuum.}
                \label{fig:Model_pheno}
\end{figure*}

Effectively, the current rise and fall through the Fano resonance can be considered as a induction current "flow" through the continuum, depending on the "flow" direction of the (spin-polarised) current, schematically illustrated in \textbf{Fig. \ref{fig:Model_pheno}}. For a chiral molecule, the dipole is dependent on the handedness (\textbf{Fig. \ref{fig:Model_pheno}A, B}). The total electric field across the molecule is sum of the dipole/polarisation and the externally applied electric field.  Adsorption on the molecules on a metallic surface will induce screening and present an overall nett dipole/across the molecular layer, which differs for both enantiomers (\textbf{Fig. \ref{fig:Model_pheno}C, D}). Delocalisation of the electrons and their aligned spins on the redox sites will lead to an induction current upon spin-order loss (\textbf{Fig. \ref{fig:Model_pheno}E, F}). The magnitude of the inductive current is related to the assymetry of the Fano resonance, and the polarisation vector either additive or substractive to the inductive current direction. Consecutive Fano resonances (\textbf{Fig. \ref{fig:Model_pheno}G}) will push the current density either up or down and this is observed in experiments as an change in resistance of the molecular aggregate. The presence and the role of a Fano resonant state, a localised electron interfering with a continuum of electrons, is crucial for boosting the spin-induced charge transport in the CBI-GCH system and possibly is more general for redox group containing chiral supramolecular polymers.

\bibliography{references.bib}

\providecommand{\latin}[1]{#1}
\makeatletter
\providecommand{\doi}
  {\begingroup\let\do\@makeother\dospecials
  \catcode`\{=1 \catcode`\}=2 \doi@aux}
\providecommand{\doi@aux}[1]{\endgroup\texttt{#1}}
\makeatother
\providecommand*\mcitethebibliography{\thebibliography}
\csname @ifundefined\endcsname{endmcitethebibliography}
  {\let\endmcitethebibliography\endthebibliography}{}
\begin{mcitethebibliography}{47}
\providecommand*\natexlab[1]{#1}
\providecommand*\mciteSetBstSublistMode[1]{}
\providecommand*\mciteSetBstMaxWidthForm[2]{}
\providecommand*\mciteBstWouldAddEndPuncttrue
  {\def\EndOfBibitem{\unskip.}}
\providecommand*\mciteBstWouldAddEndPunctfalse
  {\let\EndOfBibitem\relax}
\providecommand*\mciteSetBstMidEndSepPunct[3]{}
\providecommand*\mciteSetBstSublistLabelBeginEnd[3]{}
\providecommand*\EndOfBibitem{}
\mciteSetBstSublistMode{f}
\mciteSetBstMaxWidthForm{subitem}{(\alph{mcitesubitemcount})}
\mciteSetBstSublistLabelBeginEnd
  {\mcitemaxwidthsubitemform\space}
  {\relax}
  {\relax}

\bibitem[Waldeck \latin{et~al.}(2021)Waldeck, Naaman, and
  Paltiel]{Waldeck2021TheMaterials}
Waldeck,~D.~H.; Naaman,~R.; Paltiel,~Y. {The spin selectivity effect in chiral
  materials}. \emph{APL Materials} \textbf{2021}, \emph{9}\relax
\mciteBstWouldAddEndPuncttrue
\mciteSetBstMidEndSepPunct{\mcitedefaultmidpunct}
{\mcitedefaultendpunct}{\mcitedefaultseppunct}\relax
\EndOfBibitem
\bibitem[Naaman \latin{et~al.}(2022)Naaman, Waldeck, and
  Fransson]{Naaman2022NewMolecules}
Naaman,~R.; Waldeck,~D.~H.; Fransson,~J. {New Perspective on Electron Transfer
  through Molecules}. \emph{Journal of Physical Chemistry Letters}
  \textbf{2022}, \emph{13}, 11753--11759\relax
\mciteBstWouldAddEndPuncttrue
\mciteSetBstMidEndSepPunct{\mcitedefaultmidpunct}
{\mcitedefaultendpunct}{\mcitedefaultseppunct}\relax
\EndOfBibitem
\bibitem[Kulkarni \latin{et~al.}(2020)Kulkarni, Mondal, Das, Grinbom,
  Tassinari, Mabesoone, Meijer, and Naaman]{Kulkarni2020a}
Kulkarni,~C.; Mondal,~A.~K.; Das,~T.~K.; Grinbom,~G.; Tassinari,~F.;
  Mabesoone,~M. F.~J.; Meijer,~E.~W.; Naaman,~R. {Highly Efficient and Tunable
  Filtering of Electrons' Spin by Supramolecular Chirality of Nanofiber‐Based
  Materials}. \emph{Advanced Materials} \textbf{2020}, \emph{32}, 1904965\relax
\mciteBstWouldAddEndPuncttrue
\mciteSetBstMidEndSepPunct{\mcitedefaultmidpunct}
{\mcitedefaultendpunct}{\mcitedefaultseppunct}\relax
\EndOfBibitem
\bibitem[G{\"{o}}hler \latin{et~al.}(2011)G{\"{o}}hler, Hamelbeck, Markus,
  Kettner, Hanne, Vager, Naaman, and Zacharias]{Gohler2011SpinDNA}
G{\"{o}}hler,~B.; Hamelbeck,~V.; Markus,~T.~Z.; Kettner,~M.; Hanne,~G.~F.;
  Vager,~Z.; Naaman,~R.; Zacharias,~H. {Spin selectivity in electron
  transmission through self-assembled monolayers of double-stranded DNA}.
  \emph{Science} \textbf{2011}, \emph{331}, 894--897\relax
\mciteBstWouldAddEndPuncttrue
\mciteSetBstMidEndSepPunct{\mcitedefaultmidpunct}
{\mcitedefaultendpunct}{\mcitedefaultseppunct}\relax
\EndOfBibitem
\bibitem[Sang \latin{et~al.}(2021)Sang, Mishra, Tassinari, Karuppannan,
  Carmieli, Teo, Migliore, Beratan, Gray, Pecht, Fransson, Waldeck, and
  Naaman]{Sang2021TemperatureAzurin}
Sang,~Y.; Mishra,~S.; Tassinari,~F.; Karuppannan,~S.~K.; Carmieli,~R.;
  Teo,~R.~D.; Migliore,~A.; Beratan,~D.~N.; Gray,~H.~B.; Pecht,~I.;
  Fransson,~J.; Waldeck,~D.~H.; Naaman,~R. {Temperature Dependence of Charge
  and Spin Transfer in Azurin}. \emph{Journal of Physical Chemistry C}
  \textbf{2021}, \emph{125}, 9875--9883\relax
\mciteBstWouldAddEndPuncttrue
\mciteSetBstMidEndSepPunct{\mcitedefaultmidpunct}
{\mcitedefaultendpunct}{\mcitedefaultseppunct}\relax
\EndOfBibitem
\bibitem[Michaeli \latin{et~al.}(2017)Michaeli, Varade, Naaman, and
  Waldeck]{Michaeli2017AMagnets}
Michaeli,~K.; Varade,~V.; Naaman,~R.; Waldeck,~D.~H. {A new approach towards
  spintronics-spintronics with no magnets}. \emph{Journal of Physics Condensed
  Matter} \textbf{2017}, \emph{29}\relax
\mciteBstWouldAddEndPuncttrue
\mciteSetBstMidEndSepPunct{\mcitedefaultmidpunct}
{\mcitedefaultendpunct}{\mcitedefaultseppunct}\relax
\EndOfBibitem
\bibitem[Mondal \latin{et~al.}(2021)Mondal, Preuss, Sleczkowski, Das, Vantomme,
  Meijer, and Naaman]{Mondal2021SpinSolvents}
Mondal,~A.~K.; Preuss,~M.~D.; Sleczkowski,~M.~L.; Das,~T.~K.; Vantomme,~G.;
  Meijer,~E.~W.; Naaman,~R. {Spin Filtering in Supramolecular Polymers
  Assembled from Achiral Monomers Mediated by Chiral Solvents}. \emph{Journal
  of the American Chemical Society} \textbf{2021}, \emph{143}, 7189--7195\relax
\mciteBstWouldAddEndPuncttrue
\mciteSetBstMidEndSepPunct{\mcitedefaultmidpunct}
{\mcitedefaultendpunct}{\mcitedefaultseppunct}\relax
\EndOfBibitem
\bibitem[Lu \latin{et~al.}(2019)Lu, Wang, Xiao, Pan, Chen, Brunecky, Berry,
  Zhu, Beard, and Vardeny]{Lu2019Spin-dependentPerovskites}
Lu,~H.; Wang,~J.; Xiao,~C.; Pan,~X.; Chen,~X.; Brunecky,~R.; Berry,~J.~J.;
  Zhu,~K.; Beard,~M.~C.; Vardeny,~Z.~V. {Spin-dependent charge transport
  through 2D chiral hybrid lead-iodide perovskites}. \emph{Science Advances}
  \textbf{2019}, \emph{5}, 1--8\relax
\mciteBstWouldAddEndPuncttrue
\mciteSetBstMidEndSepPunct{\mcitedefaultmidpunct}
{\mcitedefaultendpunct}{\mcitedefaultseppunct}\relax
\EndOfBibitem
\bibitem[Bhowmick \latin{et~al.}(2022)Bhowmick, Das, Santra, Mondal, Tassinari,
  Schwarz, Diesendruck, and
  Naaman]{Bhowmick2022Spin-inducedElectropolymerization}
Bhowmick,~D.~K.; Das,~T.~K.; Santra,~K.; Mondal,~A.~K.; Tassinari,~F.;
  Schwarz,~R.; Diesendruck,~C.~E.; Naaman,~R. {Spin-induced asymmetry reaction
  - The formation of asymmetric carbon by electropolymerization}. \emph{Science
  Advances} \textbf{2022}, \emph{8}, 1--8\relax
\mciteBstWouldAddEndPuncttrue
\mciteSetBstMidEndSepPunct{\mcitedefaultmidpunct}
{\mcitedefaultendpunct}{\mcitedefaultseppunct}\relax
\EndOfBibitem
\bibitem[Nozaki \latin{et~al.}(2013)Nozaki, Sevin{\c{c}}li, Avdoshenko,
  Gutierrez, and Cuniberti]{Nozaki2013AJunctions}
Nozaki,~D.; Sevin{\c{c}}li,~H.; Avdoshenko,~S.~M.; Gutierrez,~R.; Cuniberti,~G.
  {A parabolic model to control quantum interference in T-shaped molecular
  junctions}. \emph{Physical Chemistry Chemical Physics} \textbf{2013},
  \emph{15}, 13951--13958\relax
\mciteBstWouldAddEndPuncttrue
\mciteSetBstMidEndSepPunct{\mcitedefaultmidpunct}
{\mcitedefaultendpunct}{\mcitedefaultseppunct}\relax
\EndOfBibitem
\bibitem[Xu \latin{et~al.}(2005)Xu, Xiao, Yang, Zang, and
  Tao]{Xu2005LargeTransistor}
Xu,~B.; Xiao,~X.; Yang,~X.; Zang,~L.; Tao,~N. {Large gate modulation in the
  current of a room temperature single molecule transistor}. \emph{Journal of
  the American Chemical Society} \textbf{2005}, \emph{127}, 2386--2387\relax
\mciteBstWouldAddEndPuncttrue
\mciteSetBstMidEndSepPunct{\mcitedefaultmidpunct}
{\mcitedefaultendpunct}{\mcitedefaultseppunct}\relax
\EndOfBibitem
\bibitem[Chen \latin{et~al.}(2007)Chen, Hihath, Huang, Li, and
  Tao]{Chen2007MeasurementConductance}
Chen,~F.; Hihath,~J.; Huang,~Z.; Li,~X.; Tao,~N.~J. {Measurement of
  single-molecule conductance}. \emph{Annual Review of Physical Chemistry}
  \textbf{2007}, \emph{58}, 535--564\relax
\mciteBstWouldAddEndPuncttrue
\mciteSetBstMidEndSepPunct{\mcitedefaultmidpunct}
{\mcitedefaultendpunct}{\mcitedefaultseppunct}\relax
\EndOfBibitem
\bibitem[McCreery(2009)]{McCreery2009ElectronJunctions}
McCreery,~R.~L. {Electron transport and redox reactions in molecular electronic
  junctions}. \emph{ChemPhysChem} \textbf{2009}, \emph{10}, 2387--2391\relax
\mciteBstWouldAddEndPuncttrue
\mciteSetBstMidEndSepPunct{\mcitedefaultmidpunct}
{\mcitedefaultendpunct}{\mcitedefaultseppunct}\relax
\EndOfBibitem
\bibitem[He \latin{et~al.}(2003)He, Li, Tao, Nagahara, Amlani, and
  Tsui]{He2003DiscreteWires}
He,~X.; Li,~L.; Tao,~J.; Nagahara,~A.; Amlani,~I.; Tsui,~R. {Discrete
  conductance switching in conducting polymer wires}. \emph{Physical Review B -
  Condensed Matter and Materials Physics} \textbf{2003}, \emph{68}, 1--6\relax
\mciteBstWouldAddEndPuncttrue
\mciteSetBstMidEndSepPunct{\mcitedefaultmidpunct}
{\mcitedefaultendpunct}{\mcitedefaultseppunct}\relax
\EndOfBibitem
\bibitem[Jia \latin{et~al.}(2020)Jia, Grace, Wang, Almeshal, Huang, Wang, Chen,
  Wang, Zhou, Feng, Zhao, Huang, Lambert, and Duan]{Jia2020RedoxJunctions}
Jia,~C.; Grace,~I.~M.; Wang,~P.; Almeshal,~A.; Huang,~Z.; Wang,~Y.; Chen,~P.;
  Wang,~L.; Zhou,~J.; Feng,~Z.; Zhao,~Z.; Huang,~Y.; Lambert,~C.~J.; Duan,~X.
  {Redox Control of Charge Transport in Vertical Ferrocene Molecular Tunnel
  Junctions}. \emph{Chem} \textbf{2020}, \emph{6}, 1172--1182\relax
\mciteBstWouldAddEndPuncttrue
\mciteSetBstMidEndSepPunct{\mcitedefaultmidpunct}
{\mcitedefaultendpunct}{\mcitedefaultseppunct}\relax
\EndOfBibitem
\bibitem[Bohl \latin{et~al.}(2017)Bohl, Mignolet, Johansson, Remacle, and
  Campbell]{Bohl2017Low-lyingAlkanes}
Bohl,~E.; Mignolet,~B.; Johansson,~J.~O.; Remacle,~F.; Campbell,~E.~E.
  {Low-lying, Rydberg states of polycyclic aromatic hydrocarbons (PAHs) and
  cyclic alkanes}. \emph{Physical Chemistry Chemical Physics} \textbf{2017},
  \emph{19}, 24090--24099\relax
\mciteBstWouldAddEndPuncttrue
\mciteSetBstMidEndSepPunct{\mcitedefaultmidpunct}
{\mcitedefaultendpunct}{\mcitedefaultseppunct}\relax
\EndOfBibitem
\bibitem[Migliore and Nitzan(2013)Migliore, and
  Nitzan]{Migliore2013IrreversibilityJunctions}
Migliore,~A.; Nitzan,~A. {Irreversibility and hysteresis in redox molecular
  conduction junctions}. \emph{Journal of the American Chemical Society}
  \textbf{2013}, \emph{135}, 9420--9432\relax
\mciteBstWouldAddEndPuncttrue
\mciteSetBstMidEndSepPunct{\mcitedefaultmidpunct}
{\mcitedefaultendpunct}{\mcitedefaultseppunct}\relax
\EndOfBibitem
\bibitem[Li \latin{et~al.}(2015)Li, Baghernejad, Qusiy, Zsolt~Manrique, Zhang,
  Hamill, Fu, Broekmann, Hong, Wandlowski, Zhang, and
  Lambert]{Li2015Three-StateTuning}
Li,~Y.; Baghernejad,~M.; Qusiy,~A.~G.; Zsolt~Manrique,~D.; Zhang,~G.;
  Hamill,~J.; Fu,~Y.; Broekmann,~P.; Hong,~W.; Wandlowski,~T.; Zhang,~D.;
  Lambert,~C. {Three-State Single-Molecule Naphthalenediimide Switch:
  Integration of a Pendant Redox Unit for Conductance Tuning}. \emph{Angewandte
  Chemie - International Edition} \textbf{2015}, \emph{54}, 13586--13589\relax
\mciteBstWouldAddEndPuncttrue
\mciteSetBstMidEndSepPunct{\mcitedefaultmidpunct}
{\mcitedefaultendpunct}{\mcitedefaultseppunct}\relax
\EndOfBibitem
\bibitem[Anderson(1966)]{Anderson1966TheoryMetals}
Anderson,~P.~W. {Theory of localized magnetic states in metals}. \emph{Journal
  of Applied Physics} \textbf{1966}, \emph{37}, 1194\relax
\mciteBstWouldAddEndPuncttrue
\mciteSetBstMidEndSepPunct{\mcitedefaultmidpunct}
{\mcitedefaultendpunct}{\mcitedefaultseppunct}\relax
\EndOfBibitem
\bibitem[Kulkarni \latin{et~al.}(2013)Kulkarni, Munirathinam, and
  George]{Kulkarni2013Self-assemblyAmplification}
Kulkarni,~C.; Munirathinam,~R.; George,~S.~J. {Self-assembly of coronene
  bisimides: Mechanistic insight and chiral amplification}. \emph{Chemistry - A
  European Journal} \textbf{2013}, \emph{19}, 11270--11278\relax
\mciteBstWouldAddEndPuncttrue
\mciteSetBstMidEndSepPunct{\mcitedefaultmidpunct}
{\mcitedefaultendpunct}{\mcitedefaultseppunct}\relax
\EndOfBibitem
\bibitem[Sang \latin{et~al.}(2022)Sang, Tassinari, Santra, Zhang, Fontanesi,
  Bloom, Waldeck, Fransson, and Naaman]{Sang2022ChiralityReduction}
Sang,~Y.; Tassinari,~F.; Santra,~K.; Zhang,~W.; Fontanesi,~C.; Bloom,~B.~P.;
  Waldeck,~D.~H.; Fransson,~J.; Naaman,~R. {Chirality enhances oxygen
  reduction}. \emph{Proceedings of the National Academy of Sciences of the
  United States of America} \textbf{2022}, \emph{119}, 1--8\relax
\mciteBstWouldAddEndPuncttrue
\mciteSetBstMidEndSepPunct{\mcitedefaultmidpunct}
{\mcitedefaultendpunct}{\mcitedefaultseppunct}\relax
\EndOfBibitem
\bibitem[Liang \latin{et~al.}(2022)Liang, Banjac, Martin, Zigon, Lee,
  Vanthuyne, Garc{\'{e}}s-Pineda, Gal{\'{a}}n-Mascar{\'{o}}s, Hu, Avarvari, and
  Lingenfelder]{Liang2022EnhancementElectrodes}
Liang,~Y.; Banjac,~K.; Martin,~K.; Zigon,~N.; Lee,~S.; Vanthuyne,~N.;
  Garc{\'{e}}s-Pineda,~F.~A.; Gal{\'{a}}n-Mascar{\'{o}}s,~J.~R.; Hu,~X.;
  Avarvari,~N.; Lingenfelder,~M. {Enhancement of electrocatalytic oxygen
  evolution by chiral molecular functionalization of hybrid 2D electrodes}.
  \emph{Nature Communications} \textbf{2022}, \emph{13}, 1--9\relax
\mciteBstWouldAddEndPuncttrue
\mciteSetBstMidEndSepPunct{\mcitedefaultmidpunct}
{\mcitedefaultendpunct}{\mcitedefaultseppunct}\relax
\EndOfBibitem
\bibitem[Mtangi \latin{et~al.}(2017)Mtangi, Tassinari, Vankayala,
  Vargas~Jentzsch, Adelizzi, Palmans, Fontanesi, Meijer, and
  Naaman]{Mtangi2017ControlSplitting}
Mtangi,~W.; Tassinari,~F.; Vankayala,~K.; Vargas~Jentzsch,~A.; Adelizzi,~B.;
  Palmans,~A.~R.; Fontanesi,~C.; Meijer,~E.~W.; Naaman,~R. {Control of
  Electrons' Spin Eliminates Hydrogen Peroxide Formation during Water
  Splitting}. \emph{Journal of the American Chemical Society} \textbf{2017},
  \emph{139}, 2794--2798\relax
\mciteBstWouldAddEndPuncttrue
\mciteSetBstMidEndSepPunct{\mcitedefaultmidpunct}
{\mcitedefaultendpunct}{\mcitedefaultseppunct}\relax
\EndOfBibitem
\bibitem[Urdampilleta \latin{et~al.}(2011)Urdampilleta, Klyatskaya, Cleuziou,
  Ruben, and Wernsdorfer]{Urdampilleta2011SupramolecularValves}
Urdampilleta,~M.; Klyatskaya,~S.; Cleuziou,~J.~P.; Ruben,~M.; Wernsdorfer,~W.
  {Supramolecular spin valves}. \emph{Nature Materials} \textbf{2011},
  \emph{10}, 502--506\relax
\mciteBstWouldAddEndPuncttrue
\mciteSetBstMidEndSepPunct{\mcitedefaultmidpunct}
{\mcitedefaultendpunct}{\mcitedefaultseppunct}\relax
\EndOfBibitem
\bibitem[Hong and Kim(2013)Hong, and Kim]{Hong2013Fano-resonance-drivenMagnets}
Hong,~K.; Kim,~W.~Y. {Fano-resonance-driven spin-valve effect using
  single-molecule magnets}. \emph{Angewandte Chemie - International Edition}
  \textbf{2013}, \emph{52}, 3389--3393\relax
\mciteBstWouldAddEndPuncttrue
\mciteSetBstMidEndSepPunct{\mcitedefaultmidpunct}
{\mcitedefaultendpunct}{\mcitedefaultseppunct}\relax
\EndOfBibitem
\bibitem[Nguyen \latin{et~al.}(2022)Nguyen, Rasabathina, Hellwig, Sharma,
  Salvan, Yochelis, Paltiel, Baczewski, and
  Tegenkamp]{Nguyen2022CooperativeMicroscopy}
Nguyen,~T. N.~H.; Rasabathina,~L.; Hellwig,~O.; Sharma,~A.; Salvan,~G.;
  Yochelis,~S.; Paltiel,~Y.; Baczewski,~L.~T.; Tegenkamp,~C. {Cooperative
  Effect of Electron Spin Polarization in Chiral Molecules Studied with
  Non-Spin-Polarized Scanning Tunneling Microscopy}. \emph{ACS Applied
  Materials and Interfaces} \textbf{2022}, \relax
\mciteBstWouldAddEndPunctfalse
\mciteSetBstMidEndSepPunct{\mcitedefaultmidpunct}
{}{\mcitedefaultseppunct}\relax
\EndOfBibitem
\bibitem[Ko \latin{et~al.}(2022)Ko, Zhu, Tassinari, Bullard, Zhang, Beratan,
  Naaman, and Therien]{Ko2022TwistedTemperature}
Ko,~C.~H.; Zhu,~Q.; Tassinari,~F.; Bullard,~G.; Zhang,~P.; Beratan,~D.~N.;
  Naaman,~R.; Therien,~M.~J. {Twisted molecular wires polarize spin currents at
  room temperature}. \emph{Proceedings of the National Academy of Sciences of
  the United States of America} \textbf{2022}, \emph{119}, 1--7\relax
\mciteBstWouldAddEndPuncttrue
\mciteSetBstMidEndSepPunct{\mcitedefaultmidpunct}
{\mcitedefaultendpunct}{\mcitedefaultseppunct}\relax
\EndOfBibitem
\bibitem[Mishra \latin{et~al.}(2020)Mishra, Mondal, Smolinsky, Naaman, Maeda,
  Nishimura, Taniguchi, Yoshida, Takayama, and Yashima]{Mishra2020SpinPolymers}
Mishra,~S.; Mondal,~A.~K.; Smolinsky,~E.~Z.; Naaman,~R.; Maeda,~K.;
  Nishimura,~T.; Taniguchi,~T.; Yoshida,~T.; Takayama,~K.; Yashima,~E. {Spin
  Filtering Along Chiral Polymers}. \emph{Angewandte Chemie - International
  Edition} \textbf{2020}, \emph{59}, 14671--14676\relax
\mciteBstWouldAddEndPuncttrue
\mciteSetBstMidEndSepPunct{\mcitedefaultmidpunct}
{\mcitedefaultendpunct}{\mcitedefaultseppunct}\relax
\EndOfBibitem
\bibitem[Mondal \latin{et~al.}(2020)Mondal, Brown, Mishra, Makam, Wing, Gilead,
  Wiesenfeld, Leitus, Shimon, Carmieli, Ehre, Kamieniarz, Fransson, Hod,
  Kronik, Gazit, and Naaman]{Mondal2020Long-RangeMagnetization}
Mondal,~A.~K. \latin{et~al.}  {Long-Range Spin-Selective Transport in Chiral
  Metal-Organic Crystals with Temperature-Activated Magnetization}. \emph{ACS
  Nano} \textbf{2020}, \emph{14}, 16624--16633\relax
\mciteBstWouldAddEndPuncttrue
\mciteSetBstMidEndSepPunct{\mcitedefaultmidpunct}
{\mcitedefaultendpunct}{\mcitedefaultseppunct}\relax
\EndOfBibitem
\bibitem[Lu \latin{et~al.}(2021)Lu, Wang, He, Zhou, Yang, Wang, Cao, He, Pan,
  Yang, and Song]{Lu2021HighlyHalides}
Lu,~Y.; Wang,~Q.; He,~R.; Zhou,~F.; Yang,~X.; Wang,~D.; Cao,~H.; He,~W.;
  Pan,~F.; Yang,~Z.; Song,~C. {Highly Efficient Spin-Filtering Transport in
  Chiral Hybrid Copper Halides}. \emph{Angewandte Chemie - International
  Edition} \textbf{2021}, \emph{60}, 23578--23583\relax
\mciteBstWouldAddEndPuncttrue
\mciteSetBstMidEndSepPunct{\mcitedefaultmidpunct}
{\mcitedefaultendpunct}{\mcitedefaultseppunct}\relax
\EndOfBibitem
\bibitem[Yuan \latin{et~al.}(2019)Yuan, Ji, Xing, Li, Gazit, and
  Yan]{Yuan2019HierarchicallyCrystals}
Yuan,~C.; Ji,~W.; Xing,~R.; Li,~J.; Gazit,~E.; Yan,~X. {Hierarchically oriented
  organization in supramolecular peptide crystals}. \emph{Nature Reviews
  Chemistry} \textbf{2019}, \emph{3}, 567--588\relax
\mciteBstWouldAddEndPuncttrue
\mciteSetBstMidEndSepPunct{\mcitedefaultmidpunct}
{\mcitedefaultendpunct}{\mcitedefaultseppunct}\relax
\EndOfBibitem
\bibitem[Potticary \latin{et~al.}(2016)Potticary, Terry, Bell,
  Papanikolopoulos, Christianen, Engelkamp, Collins, Fontanesi,
  Kociok-K{\"{o}}hn, Crampin, Da~Como, and Hall]{Potticary2016AnGrowth}
Potticary,~J.; Terry,~L.~R.; Bell,~C.; Papanikolopoulos,~A.~N.;
  Christianen,~P.~C.; Engelkamp,~H.; Collins,~A.~M.; Fontanesi,~C.;
  Kociok-K{\"{o}}hn,~G.; Crampin,~S.; Da~Como,~E.; Hall,~S.~R. {An unforeseen
  polymorph of coronene by the application of magnetic fields during crystal
  growth}. \emph{Nature Communications} \textbf{2016}, \emph{7}, 1--7\relax
\mciteBstWouldAddEndPuncttrue
\mciteSetBstMidEndSepPunct{\mcitedefaultmidpunct}
{\mcitedefaultendpunct}{\mcitedefaultseppunct}\relax
\EndOfBibitem
\bibitem[Kulkarni \latin{et~al.}(2017)Kulkarni, Korevaar, Bejagam, Palmans,
  Meijer, and George]{Kulkarni2017SolventPockets}
Kulkarni,~C.; Korevaar,~P.~A.; Bejagam,~K.~K.; Palmans,~A.~R.; Meijer,~E.~W.;
  George,~S.~J. {Solvent clathrate driven dynamic stereomutation of a
  supramolecular polymer with molecular pockets}. \emph{Journal of the American
  Chemical Society} \textbf{2017}, \emph{139}, 13867--13875\relax
\mciteBstWouldAddEndPuncttrue
\mciteSetBstMidEndSepPunct{\mcitedefaultmidpunct}
{\mcitedefaultendpunct}{\mcitedefaultseppunct}\relax
\EndOfBibitem
\bibitem[Daub \latin{et~al.}(2021)Daub, Janssen, and
  Hendriks]{Daub2021Imide-BasedBatteries}
Daub,~N.; Janssen,~R.~A.; Hendriks,~K.~H. {Imide-Based Multielectron Anolytes
  as High-Performance Materials in Nonaqueous Redox Flow Batteries}. \emph{ACS
  Applied Energy Materials} \textbf{2021}, \emph{4}, 9248--9257\relax
\mciteBstWouldAddEndPuncttrue
\mciteSetBstMidEndSepPunct{\mcitedefaultmidpunct}
{\mcitedefaultendpunct}{\mcitedefaultseppunct}\relax
\EndOfBibitem
\bibitem[Ghosh \latin{et~al.}(2020)Ghosh, Mishra, Avigad, Bloom, Baczewski,
  Yochelis, Paltiel, Naaman, and Waldeck]{Ghosh2020EffectInterfaces}
Ghosh,~S.; Mishra,~S.; Avigad,~E.; Bloom,~B.~P.; Baczewski,~L.~T.;
  Yochelis,~S.; Paltiel,~Y.; Naaman,~R.; Waldeck,~D.~H. {Effect of Chiral
  Molecules on the Electron's Spin Wavefunction at Interfaces}. \emph{Journal
  of Physical Chemistry Letters} \textbf{2020}, \emph{11}, 1550--1557\relax
\mciteBstWouldAddEndPuncttrue
\mciteSetBstMidEndSepPunct{\mcitedefaultmidpunct}
{\mcitedefaultendpunct}{\mcitedefaultseppunct}\relax
\EndOfBibitem
\bibitem[Eckshtain-Levi \latin{et~al.}(2016)Eckshtain-Levi, Capua,
  Refaely-Abramson, Sarkar, Gavrilov, Mathew, Paltiel, Levy, Kronik, and
  Naaman]{Eckshtain-Levi2016ColdMonolayers}
Eckshtain-Levi,~M.; Capua,~E.; Refaely-Abramson,~S.; Sarkar,~S.; Gavrilov,~Y.;
  Mathew,~S.~P.; Paltiel,~Y.; Levy,~Y.; Kronik,~L.; Naaman,~R. {Cold
  denaturation induces inversion of dipole and spin transfer in chiral peptide
  monolayers}. \emph{Nature Communications} \textbf{2016}, \emph{7}, 1--9\relax
\mciteBstWouldAddEndPuncttrue
\mciteSetBstMidEndSepPunct{\mcitedefaultmidpunct}
{\mcitedefaultendpunct}{\mcitedefaultseppunct}\relax
\EndOfBibitem
\bibitem[Alwan and Dubi(2021)Alwan, and Dubi]{Alwan2021SpinterfaceEffect}
Alwan,~S.; Dubi,~Y. {Spinterface Origin for the Chirality-Induced
  Spin-Selectivity Effect}. \emph{Journal of the American Chemical Society}
  \textbf{2021}, \emph{143}, 14235--14241\relax
\mciteBstWouldAddEndPuncttrue
\mciteSetBstMidEndSepPunct{\mcitedefaultmidpunct}
{\mcitedefaultendpunct}{\mcitedefaultseppunct}\relax
\EndOfBibitem
\bibitem[Ortiz(2020)]{Ortiz2020Dyson-orbitalMolecules}
Ortiz,~J.~V. {Dyson-orbital concepts for description of electrons in
  molecules}. \emph{Journal of Chemical Physics} \textbf{2020},
  \emph{153}\relax
\mciteBstWouldAddEndPuncttrue
\mciteSetBstMidEndSepPunct{\mcitedefaultmidpunct}
{\mcitedefaultendpunct}{\mcitedefaultseppunct}\relax
\EndOfBibitem
\bibitem[Pastukhova \latin{et~al.}(2017)Pastukhova, Samos, Zoppi, Pavlica,
  Mathew, Bratina, Siegel, and Baldridge]{Pastukhova2017EvidenceFilms}
Pastukhova,~N.; Samos,~L.~M.; Zoppi,~L.; Pavlica,~E.; Mathew,~J.; Bratina,~G.;
  Siegel,~J.~S.; Baldridge,~K.~K. {Evidence of enhanced photocurrent response
  in corannulene films}. \emph{RSC Advances} \textbf{2017}, \emph{7},
  45601--45606\relax
\mciteBstWouldAddEndPuncttrue
\mciteSetBstMidEndSepPunct{\mcitedefaultmidpunct}
{\mcitedefaultendpunct}{\mcitedefaultseppunct}\relax
\EndOfBibitem
\bibitem[Feng \latin{et~al.}(2008)Feng, Zhao, and Petek]{Feng2008Atomlike60}
Feng,~M.; Zhao,~J.; Petek,~H. {Atomlike, Hollow-Core – Bound Molecular
  Orbitals of C 60}. \textbf{2008}, \emph{320}, 359--363\relax
\mciteBstWouldAddEndPuncttrue
\mciteSetBstMidEndSepPunct{\mcitedefaultmidpunct}
{\mcitedefaultendpunct}{\mcitedefaultseppunct}\relax
\EndOfBibitem
\bibitem[Wang \latin{et~al.}(2021)Wang, Mujica, and
  Lai]{Wang2021SpinImplications}
Wang,~C.~Z.; Mujica,~V.; Lai,~Y.~C. {Spin Fano Resonances in Chiral Molecules:
  An Alternative Mechanism for the CISS Effect and Experimental Implications}.
  \emph{Nano Letters} \textbf{2021}, \emph{21}, 10423--10430\relax
\mciteBstWouldAddEndPuncttrue
\mciteSetBstMidEndSepPunct{\mcitedefaultmidpunct}
{\mcitedefaultendpunct}{\mcitedefaultseppunct}\relax
\EndOfBibitem
\bibitem[Van~Bree \latin{et~al.}(2014)Van~Bree, Silov, Koenraad, and
  Flatt{\'{e}}]{VanBree2014Spin-orbit-inducedNanostructure}
Van~Bree,~J.; Silov,~A.~Y.; Koenraad,~P.~M.; Flatt{\'{e}},~M.~E.
  {Spin-orbit-induced circulating currents in a semiconductor nanostructure}.
  \emph{Physical Review Letters} \textbf{2014}, \emph{112}, 1--5\relax
\mciteBstWouldAddEndPuncttrue
\mciteSetBstMidEndSepPunct{\mcitedefaultmidpunct}
{\mcitedefaultendpunct}{\mcitedefaultseppunct}\relax
\EndOfBibitem
\bibitem[Rodrigues~da Cruz and Flatt{\'{e}}(2022)Rodrigues~da Cruz, and
  Flatt{\'{e}}]{RodriguesdaCruz2022DissipationlessGas}
Rodrigues~da Cruz,~A.; Flatt{\'{e}},~M.~E. {Dissipationless circulating
  currents and fringe magnetic fields near a single spin embedded in a
  two-dimensional electron gas}. \textbf{2022}, 78\relax
\mciteBstWouldAddEndPuncttrue
\mciteSetBstMidEndSepPunct{\mcitedefaultmidpunct}
{\mcitedefaultendpunct}{\mcitedefaultseppunct}\relax
\EndOfBibitem
\bibitem[Sader \latin{et~al.}(1995)Sader, Larson, Mulvaney, and
  White]{Sader1995MethodCantilevers}
Sader,~J.~E.; Larson,~I.; Mulvaney,~P.; White,~L.~R. {Method for the
  calibration of atomic force microscope cantilevers}. \emph{Review of
  Scientific Instruments} \textbf{1995}, \emph{66}, 3789--3798\relax
\mciteBstWouldAddEndPuncttrue
\mciteSetBstMidEndSepPunct{\mcitedefaultmidpunct}
{\mcitedefaultendpunct}{\mcitedefaultseppunct}\relax
\EndOfBibitem
\bibitem[Jiang \latin{et~al.}(2019)Jiang, Weber, Puglisi, Pavan, Larcher,
  Frammelsberger, Benstetter, and Lanza]{Jiang2019UnderstandingMicroscopy}
Jiang,~L.; Weber,~J.; Puglisi,~F.~M.; Pavan,~P.; Larcher,~L.;
  Frammelsberger,~W.; Benstetter,~G.; Lanza,~M. {Understanding current
  instabilities in conductive atomic force microscopy}. \emph{Materials}
  \textbf{2019}, \emph{12}, 1--10\relax
\mciteBstWouldAddEndPuncttrue
\mciteSetBstMidEndSepPunct{\mcitedefaultmidpunct}
{\mcitedefaultendpunct}{\mcitedefaultseppunct}\relax
\EndOfBibitem
\bibitem[Celano(2019)]{Celano2019TheNanoelectronics}
Celano,~U. \emph{NanoScience and Technology}; 2019; pp 1--28\relax
\mciteBstWouldAddEndPuncttrue
\mciteSetBstMidEndSepPunct{\mcitedefaultmidpunct}
{\mcitedefaultendpunct}{\mcitedefaultseppunct}\relax
\EndOfBibitem
\end{mcitethebibliography}

\end{document}